\documentclass[pra,twocolumn,notitlepage,floatfix]{revtex4-2}
\usepackage[T1]{fontenc}
\usepackage[colorlinks=true,urlcolor=blue,citecolor=blue,anchorcolor=red]{hyperref}
\usepackage{graphics,url,color,physics,bbm,soul,mathtools,amssymb,amsmath,dsfont,booktabs}

\begin{document}

\title{Towards adiabatic quantum computing using compressed quantum circuits}

\author{Conor Mc Keever}
\email{conor.mckeever@quantinuum.com}
\affiliation{Quantinuum, Partnership House, Carlisle Place, London SW1P 1BX, United Kingdom}

\author{Michael Lubasch}
\affiliation{Quantinuum, Partnership House, Carlisle Place, London SW1P 1BX, United Kingdom}

\date{June 20, 2024}

\begin{abstract}
We describe tensor network algorithms to optimize quantum circuits for adiabatic quantum computing.
To suppress diabatic transitions, we include counterdiabatic driving in the optimization and utilize variational matrix product operators to represent adiabatic gauge potentials.
Traditionally, Trotter product formulas are used to turn adiabatic time evolution into quantum circuits and the addition of counterdiabatic driving increases the circuit depth per time step.
Instead, we classically optimize a parameterized quantum circuit of fixed depth to simultaneously capture adiabatic evolution together with counterdiabatic driving over many time steps.
The methods are applied to the ground state preparation of quantum Ising chains with transverse and longitudinal fields.
We show that the classically optimized circuits can significantly outperform Trotter product formulas.
Additionally, we discuss how the approach can be used for combinatorial optimization.
\end{abstract}

\maketitle

\section{Introduction}

To make use of current quantum computing technologies, adiabatic time evolution~\cite{KaNi98, FaEtAl00, AlLi18, HaEtAl20} is a promising concept that underpins, for example, the computations of the famous D-Wave device~\cite{JoEtAl11} and motivates the popular quantum approximate optimization algorithm~\cite{FaGoGu14}.
In the adiabatic approach, the quantum device prepares a trivial ground state and then realizes time evolution with a specific, time-dependent Hamiltonian, whose initial ground state is the trivial one and whose final ground state encodes the desired quantum computational result.
The procedure is successful if the evolution occurs sufficiently slowly and the total evolution time scales $\propto 1/\Delta^{2}$~\cite{Ze32} where $\Delta$ is the minimum energy difference between the ground and the first excited state during the evolution.
On digital quantum computers, time evolution can be readily realized using Trotter product formulas~\cite{Ll96}, but long evolution times can lead to deep circuits.
One powerful strategy to reduce the adiabatic evolution time is based on the inclusion of additional terms in the Hamiltonian that suppress unwanted transitions during the time evolution~\cite{DeRi03, DeRi05, Be09, ChEtAl10, DeRaZu12, SaEtAl14, ChEtAl21}, which we refer to as counterdiabatic driving~\cite{SePo17, KoEtAl17}.
These additional Hamiltonian terms, however, lead to deeper Trotter circuits per time step on a gate-based quantum device~\cite{HeEtAl21a, HeEtAl21b, HeChSo22, HeEtAl22A}.
Therefore, adiabatic time evolution over many time steps, with or without counterdiabatic driving, can still be a challenge for current digital quantum computers.

In this article, we extend the tensor network~\cite{VeMuCi08, Or14, Ba23} toolbox for parameterized quantum circuit (PQC)~\cite{BeEtAl19, CeEtAl21, BhEtAl22} optimization of Hamiltonian simulation~\cite{MaEtAl23, TeHaLu23, McLu23} to tackle adiabatic quantum computing enhanced by counterdiabatic driving.
The key ingredients of counterdiabatic driving, that suppress unwanted transitions, are the auxiliary Hamiltonian terms~\cite{DeRi03, DeRi05, Be09, ChEtAl10, DeRaZu12, SaEtAl14, ChEtAl21} which we refer to as the adiabatic gauge potential (AGP)~\cite{SePo17, KoEtAl17, Pandey2020}.
We study the suitability of a variational matrix product operator (MPO)~\cite{VeGaCi04, ZwVi04} ansatz to approximate the AGP.
The MPO ansatz naturally fits within our tensor network approach and we also show that it can have advantages over the popular nested commutator (NC) ansatz~\cite{ClEtAl19}.
Following~\cite{McLu23}, we divide the total evolution time into several chunks of shorter time intervals.
For each chunk we optimize a specific PQC, dedicated to that chunk, to capture the counterdiabatic dynamics over the shorter evolution time of the chunk.
After the classical optimization, the sequence of PQCs represents the entire counterdiabatic evolution and can be evaluated on a quantum computer.
In the context of transverse-field quantum Ising chains with a longitudinal field, we numerically demonstrate that, compared with Trotter circuits, the classically optimized PQCs can improve ground state fidelities by a factor of five and energy accuracies by a factor of three.

Our approach is inspired by several important articles.
Firstly, variational quantum algorithms~\cite{CeEtAl21, BhEtAl22, Ma23} can be used to simulate quantum dynamics~\cite{LiBe17, YuEtAl19, BeFiLu21, BaViCa21} and beautiful proposals exist that make use of counterdiabaticity by including additional gates in the PQC ansatz that come from counterdiabatic driving~\cite{ChEtAl22, HeEtAl22A, ChEtAl23a, ChEtAl23b, MaEtAl24}.
Because variational quantum algorithms can have a large quantum computational overhead, e.g.\ when small gradients need to be measured on quantum computers~\cite{McEtAl18, CeEtAl21B, Holmes2021, Holmes2022, Arrasmith2022}, we instead perform the entire PQC optimization on classical computers.
Furthermore, our PQC ansatz does not necessarily need to contain these additional counterdiabatic gates.
Provided that it remains efficient for tensor network methods, the PQC architecture can be freely chosen, e.g.\ a PQC with a quantum-hardware-efficient structure composed of hardware-native gates can be used.
Secondly, classical tensor network algorithms have already successfully simulated the entire evolution corresponding to certain adiabatic protocols~\cite{BaEtAl06, BaEtAl15, LaEtAl23, GrDr24}.
In contrast to these simulations, our procedure does not require the classical tensor network optimization to be capable of capturing the entire evolution; instead, only short chunks of the evolution need to be classically simulable.
The PQC compression of time evolution for each chunk can be performed such as to exhaust the capabilities of classical computing; then appending all PQCs can create a quantum circuit that is hard to simulate for classical computers but efficient to run on a quantum computer.

The article has the following structure.
Section~\ref{sec:Methods} contains the description of the numerical methods, including the variational optimization of the MPO gauge potential in Sec.~\ref{subsec:Methods:1} and the PQCs for the adiabatic evolution in Sec.~\ref{subsec:Methods:2}.
In Sec.~\ref{sec:Results}, we present the results of this study, i.e.\ the comparison between the nested-commutator and the MPO gauge potentials in Sec.~\ref{subsec:Results:1} as well as the comparison between the Trotter and the classically optimized circuits in Sec.~\ref{subsec:Results:2}.
In Sec.~\ref{subsec:Results:3} we compare our methods to other tensor network based techniques.
We discuss how to adapt the circuit optimization procedures for other problems, such as combinatorial optimization, in Sec.~\ref{sec:Discussion}.
Technical details are provided in the Appendixes.

\section{Methods}
\label{sec:Methods}

\begin{figure}
\centering
\includegraphics[width=86.634mm]{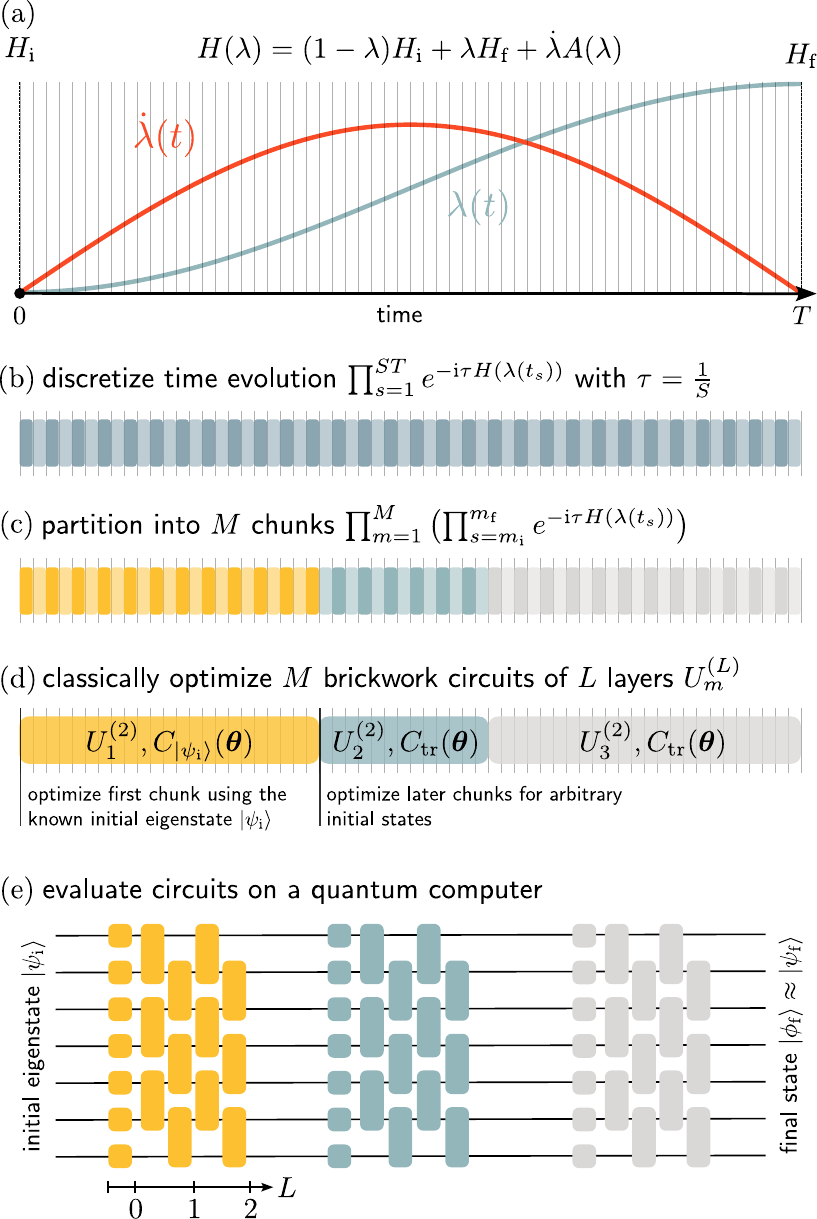}
\caption{\label{fig:1}
The proposed method to approximate counterdiabatic dynamics by a product of classically optimized quantum circuits as explained in Sec.~\ref{sec:Methods}.
This example uses $M = 3$ chunks of $L = 2$ layer brickwork parameterized quantum circuits (PQCs).
}
\end{figure}

The goal of our approach is to approximate adiabatic evolution, including counterdiabatic terms, as a quantum circuit.
The adiabatic dynamics obey a Schr\"{o}dinger equation
\begin{equation}
 \text{i} \partial_{t} |\psi\rangle = H(\lambda(t)) |\psi\rangle,
\end{equation}
where the time-dependent Hamiltonian $H(\lambda(t))$ has the form shown in Fig.~\ref{fig:1}~(a).
At the beginning, $t = 0$, $H(\lambda(0)) = H(0)$ is equivalent to the initial Hamiltonian, $H_{\text{i}}$, whose ground state is the initial eigenstate $|\psi_{\text{i}}\rangle$.
At the end, $t = T$, $H(\lambda(T)) = H(1)$ is equivalent to the final Hamiltonian, $H_{\text{f}}$, whose ground state is the final eigenstate $|\psi_{\text{f}}\rangle$.
During the evolution, $H(\lambda)$ is a linear combination of $H_{\text{i}}$ and $H_{\text{f}}$ controlled by the adiabatic parameter $\lambda(t)$ and, additionally, contains a contribution from the AGP $A(\lambda)$~\cite{SePo17, KoEtAl17} (see Appendix~\ref{app:A}) multiplied by the time derivative of the adiabatic parameter $\dot{\lambda}$.

To proceed, we first describe in Sec.~\ref{subsec:Methods:1} how to variationally approximate the AGP at any point along the adiabatic path as an MPO.
We then describe in Sec.~\ref{subsec:Methods:2} how the full counterdiabatic dynamics can be approximated using a product of classically optimized quantum circuits.

\subsection{Variational matrix product operators for adiabatic gauge potentials}
\label{subsec:Methods:1}

We obtain the gauge potential $A(\lambda)$ in Fig.~\ref{fig:1} (a) using a variational procedure~\cite{SePo17, KoEtAl17} based on an MPO ansatz that we numerically optimize to minimize a certain cost function, see Appendix~\ref{app:B}.
The MPO ansatz is characterized by the MPO bond dimension $\chi$ that determines the expressive power and computational cost related to the ansatz~\cite{VeGaCi04, ZwVi04}.
For the optimization, it is useful to write the variational gauge potential operator $\tilde{A}$ as a state $|\tilde{A}\rangle$ (by making use of the well-known Choi-Jamio\l{}kowski isomorphism).
Then the cost function can be conveniently expressed in terms of a system of linear equations
\begin{equation}\label{eq:LinEqs}
 (\mathcal{H} \otimes \mathds{1} - \mathds{1} \otimes \mathcal{H}^{T} ) |\tilde{A}\rangle = -\text{i}|\partial_{\lambda} \mathcal{H}\rangle,
\end{equation}
cf.\ Eq.~\eqref{eq:LE}, where $\mathds{1}$ denotes the identity operator, $\mathcal{H}$ is $H(\lambda)$ without the gauge potential term, i.e.\ in our case $\mathcal{H} = (1-\lambda) H_{\text{i}} + \lambda H_{\text{f}}$, and $(\cdot)^{T}$ is the transpose of $(\cdot)$.
Note that the matrix $(\mathcal{H} \otimes \mathds{1} - \mathds{1} \otimes \mathcal{H}^{T})$ is rank-deficient.
The variational optimization proceeds using tensor network algorithms for linear equations~\cite{OsDo12, LuMoJa18}.
These algorithms sweep over the MPO tensors and, for each tensor, solve a specific system of linear equations.
Because the matrices of these linear equations can be rank-deficient or ill-conditioned, we add a small regularization parameter $\eta$, i.e.\ the diagonal matrix $\eta \mathds{1}$, to them before solving the linear equations.

\subsection{Classical optimization of quantum circuits for counterdiabatic dynamics}
\label{subsec:Methods:2}

For the circuit optimization we use an iterative procedure which generalizes~\cite{McLu23} to time-dependent Hamiltonians.
The procedure is illustrated in Fig.~\ref{fig:1}~(b-d).
The total evolution over time $T$ is cut into $S T$ slices of shorter evolutions over small time steps $\tau = 1/S$ as illustrated in Fig.~\ref{fig:1}~(b).
For the evolution of each slice, we assume $\lambda(t) = \lambda(t_{s})$ where $t_{s}$ is the midpoint of slice $s$, so that during every slice evolution the Hamiltonian is time-independent.
At each time slice $s = 1, 2, \dots, S T$, the short-time evolution operator is approximated using a first-order Taylor approximation
\begin{equation}
e^{- \text{i} \tau ( \mathcal{H}_{s} + \dot{\lambda}_{s} \tilde{A}_{s})} = \mathds{1} - \text{i} \tau \mathcal{H}_{s} - \text{i} \tau \dot{\lambda}_{s} \tilde{A}_{s} + O(\tau^{2}).
\end{equation}
Here $\mathcal{H}_{s}$, $\dot{\lambda}_{s}$ and $\tilde{A}_{s}$ are equivalent to $\mathcal{H}$, $\dot{\lambda}$ and $\tilde{A}$ evaluated at time slice $s$, respectively.
The operator $W_{s} = \mathds{1} - \text{i} \tau \mathcal{H}_{s} + O(\tau^{2})$ can be represented efficiently as an MPO using the $W^{I}$ or $W^{II}$ method~\cite{ZaletelMPO} or, for higher orders of $\tau$, using the approach of~\cite{VaEtAl23}.
The operator $\tilde{A}_{s}$ is an MPO representation of the AGP that is calculated using the variational method of Sec.~\ref{subsec:Methods:1}.
Note that, for $\tilde{A}_{s}$ we can alternatively use the NC ansatz~\cite{ClEtAl19} or an expansion in terms of a Krylov basis~\cite{TaDe23}.
Also, Trotter-based approximations of the small-time evolution operator can be used where appropriate.

In order to approximate the evolution as a circuit, we split the total evolution time $T$ into $M$ chunks and assign the slices to their corresponding chunks, as shown in Fig.~\ref{fig:1}~(c).
For each of the individual chunks, we optimize a PQC $U(\boldsymbol{\theta})$ with variational parameters $\boldsymbol{\theta}$ to represent the corresponding evolution as shown in Fig.~\ref{fig:1}~(d).
The PQC corresponding to each chunk $m \in \{1, 2, \dots, M\}$ can be chosen to have any structure amenable to classical optimization.
To provide a specific example, we consider the $L$-layer brickwork circuits $U(\boldsymbol{\theta}) = U^{(L)}_{m}$ illustrated in Fig.~\ref{fig:1}~(e).
The PQC optimization then proceeds one slice after another as in~\cite{McLu23}.
More precisely, at slice $q$ of chunk $m$ of the iterative procedure, we search for the parameters $\boldsymbol{\theta} = (\theta_{1}, \theta_{2}, \dots, \theta_{K})$ of the PQC $U(\boldsymbol{\theta})$ that minimize the cost function $\| ( U(\boldsymbol{\theta}) - \exp(-\text{i} \tau H_{q}) U_{q-1} ) \rho \|^{2}_\text{F}$, where $\|\cdot\|_\text{F}$ is the Frobenius norm, $U_{q-1}$ is the optimized circuit from the previous slice and the Hermitian operator $\rho$ enables us to incorporate knowledge of the initial state.
This is equivalent to the minimization of the cost function
\begin{equation}\label{eq:CostSlice}
 C^{(q)}(\boldsymbol{\theta}) = -\Re \{ \text{tr} [U^{\dag}(\boldsymbol{\theta}) (W_{q} - \text{i} \tau \dot{\lambda}_{q} \tilde{A}_{q}) U(\boldsymbol{\theta}_{q-1}) \rho^{2} ] \},
\end{equation}
where $\Re$ denotes the real part, $\text{tr}[\cdot]$ the trace of $[\cdot]$, $(\cdot)^{\dag}$ adjoint of $(\cdot)$, and $U(\boldsymbol{\theta}_{0}) = \mathds{1}$.
If we have complete knowledge of the initial state $\ket{\psi_{\text{i}}}$, we can set $\rho = \ket{\psi_{\text{i}}}\bra{\psi_{\text{i}}}$ for the first chunk ($m = 1$); we denote the associated cost function as $C_{\ket{\psi_{\text{i}}}}(\boldsymbol{\theta})$.
In the opposite limit, where we have no knowledge of the initial state, the operator $\rho$ is set to the identity $\rho = \mathds{1}$ and we denote the associated cost function as $C_{\text{tr}}(\boldsymbol{\theta})$.
The result of the classical circuit optimization is a set of circuits $U^{(L)}_{m}(\boldsymbol{\theta})$ for $m \in \{1, 2, \dots, M\}$ which can be evaluated on a quantum computer to prepare the state $\ket{\phi_{\text{f}}} \approx \ket{\psi_{\text{f}}}$ as shown in Fig.~\ref{fig:1}~(e).

\section{Results}
\label{sec:Results}

In the following, we consider the one-dimensional nearest-neighbour quantum Ising Hamiltonian
\begin{equation}\label{eq:IsingHamiltonian}
 H_\text{Ising} = \sum_{k=1}^{N-1} J_k Z_k Z_{k+1} + \sum_{k=1}^{N} g_k X_k + \sum_{k=1}^{N} h_k Z_k.
\end{equation}
where $J_k$, $g_k$ and $h_k$ are parameters of the Hamiltonian at site $k$ and $X$ ($Z$) is the Pauli $X$ ($Z$) matrix.

\begin{figure}
\centering
\includegraphics[width=0.95\linewidth]{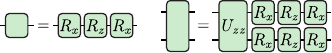}
\caption{\label{fig:2}
Parameterized quantum gates $R_{x}(\theta) = e^{-\text{i} \theta X / 2}$, $R_{z}(\theta) = e^{-\text{i} \theta Z / 2}$ and $U_{zz}(\theta) = e^{-\text{i} \theta Z \otimes Z / 2}$, where $\theta$ denotes the variational parameter, for the ansatz in Fig.~\ref{fig:1}~(e).
}
\end{figure}

In Sec.~\ref{subsec:Results:2}, we optimize PQCs $U(\boldsymbol{\theta})$ which have a brickwork structure as illustrated in Fig.~\ref{fig:1}~(e).
Each of the $M$ chunks of the brickwork circuits consists of an initial layer of single-qubit blocks followed by $L$ layers of two-qubit blocks arranged in a brickwork pattern.
The substructure of each block is shown in Fig.~\ref{fig:2}.

\subsection{Comparison of variational matrix-product-operator gauge potentials to the nested commutator approach}
\label{subsec:Results:1}

The NC method~\cite{ClEtAl19} approximates the AGP in terms of a series of nested commutators of $\mathcal{H}$ and $\partial_{\lambda}\mathcal{H}$.
The order $l$ of the NC approximation corresponds to the number of NCs retained in the series.
To facilitate a direct comparison with~\cite{ClEtAl19}, we consider an $N = 14$ qubit quantum Ising Hamiltonian~\eqref{eq:IsingHamiltonian} at one point along the adiabatic path, such that $J_k = 1.0$, and $h_k = g_k = 0.3 \lambda$ for all qubit indices $k = 1, 2, \dots, N$.
The resulting Hamiltonian is
\begin{equation}\label{eq:IsingHamiltonianNCComp}
 \mathcal{H}(\lambda) = \sum_{k=1}^{13} Z_{k}Z_{k+1} + 0.3 \lambda \sum_{k=1}^{14} \left( X_{k} + Z_{k} \right).
\end{equation}

To assess the performance of the variational MPO gauge potential $\tilde{A}$, we compute the values of the normalized cost, defined as
\begin{equation}\label{eq:NormalizedCost}
 \mathcal{C}(\tilde{A}) = \text{tr} \left[ G^{\dag}(\tilde{A}) G(\tilde{A}) \right] / \text{tr} \left[ \partial_{\lambda}\mathcal{H}^2 \right],
\end{equation}
and the normalized error
\begin{equation}\label{eq:NormalizedError}
 \mathcal{E}(\tilde{A}) = \text{tr} \left[ [G(\tilde{A}), \mathcal{H}]^{\dag} [G(\tilde{A}), \mathcal{H}] \right] / \textrm{tr} \left[ \partial_{\lambda}\mathcal{H}^2 \right]
\end{equation}
where $[O_{1}, O_{2}]$ denotes the commutator of $O_{1}$ and $O_{2}$, and we refer the reader to Eq.~\eqref{eq:G} in Appendix~\ref{app:B} for a definition of the operator $G$.

\begin{figure}
\centering
\includegraphics[width=0.95\linewidth]{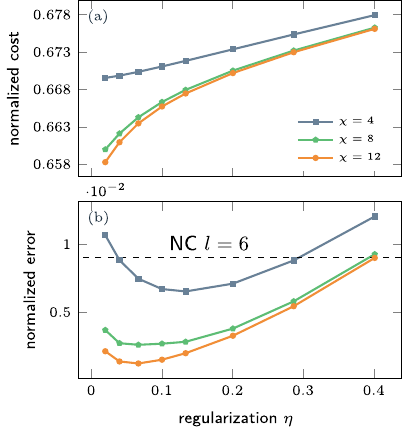}
\caption{\label{fig:3}
(a) Normalized cost $\mathcal{C}(\tilde{A})$~\eqref{eq:NormalizedCost} and (b) normalized error $\mathcal{E}(\tilde{A})$~\eqref{eq:NormalizedError} as a function of the strength of regularization used in the variational matrix product operator (MPO) method, see Sec.~\ref{subsec:Methods:1}, for different bond dimensions.
The dashed horizontal line indicates the error achieved by the nested commutator (NC) method of order $l = 6$ as reported in~\cite{ClEtAl19}.
}
\end{figure}

Results for the normalized cost and normalized error, both of which we seek to minimize during the variational optimization, are illustrated in Fig.~\ref{fig:3}.
We observe that the normalized cost in Fig.~\ref{fig:3}~(a) decreases with increasing variational MPO bond dimension $\chi$ and decreasing regularization strength $\eta$.
For the normalized error in Fig.~\ref{fig:3}~(b) we observe that, while the error decreases with increasing bond dimension $\chi$, there is an optimal regularization parameter $\eta$ which minimizes the error for each $\chi$.
Additionally, we find that the variationally optimized gauge potentials outperform the NC gauge potential of order $l = 6$ in terms of the normalized error.

\begin{figure}
\centering
\includegraphics[width=0.95\linewidth]{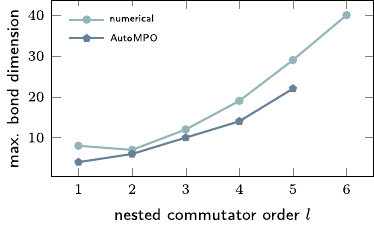}
\caption{\label{fig:4}
The maximum bond dimensions of matrix product operator (MPO) representations of the nested-commutator (NC) gauge potential of order $l$ for the $N = 14$ qubit Hamiltonian $\mathcal{H}(\lambda)$ defined in Eq.~\eqref{eq:IsingHamiltonianNCComp}.
For each NC order $l$ the prefactors $\alpha_k$ for $k = 1, 2, \dots, l$ of the NCs are calculated numerically (see~\cite{ClEtAl19}).
To construct $\tilde{A}$ as an MPO, we use numerical multiplication and addition of MPOs representing $\mathcal{H}$ and $\partial_{\lambda} \mathcal{H}$ to evaluate and sum over the NCs.
As an alternative method, we simplify the NCs algebraically using \textsf{QuantumAlgebra.jl}~\cite{QuantumAlgebra.jl} into sums of Pauli strings and construct $\tilde{A}$ using the so-called AutoMPO method~\cite{ITensor} included in the \textsf{ITensors.jl} library~\cite{ITensor-r0.3}.
}
\end{figure}

Figure~\ref{fig:4} shows the maximum bond dimension of the NC gauge potential when represented as an MPO.
We use two different techniques to construct these MPOs as described in the figure caption of Fig.~\ref{fig:4}.
When considered alongside the results of Fig.~\ref{fig:3}, we observe that the variationally optimized MPO gauge potentials outperform those constructed using the NC approach while also having a much smaller maximum bond dimension.

\subsection{Quantum circuits for counterdiabatic spectral gap traversal}
\label{subsec:Results:2}

In this section we consider an instance of the quantum Ising Hamiltonian~\eqref{eq:IsingHamiltonian} with uniform parameters $J_k = J$, $g_k = g$ and $h_k = h$ for all qubit indices $k = 1, 2, \dots, N$; we refer to this Hamiltonian as $H_\text{Ising}(J, g, h)$.
We consider an adiabatic path that starts with an initial Hamiltonian $H^\text{GT}_{\text{i}} = H_\text{Ising}(0, g^*, 0)$ and ends with a final Hamiltonian $H^\text{GT}_{\text{f}} = H_\text{Ising}(1, g^*, 1)$.
The parameter $g^*$ is numerically chosen such that the minimum spectral gap appears approximately at $\lambda = 0.5$ and the adiabatic evolution realizes a gap traversal (GT) from a paramagnetic to an antiferromagnetic phase.
Specifically, for the system sizes $N \in \{7, 15, 23, 31\}$ considered here, we choose $g^{*} \in \{0.48, 0.45, 0.435, 0.43\}$.
The resulting adiabatic Hamiltonian that interpolates between $H^\text{GT}_{\text{i}}$ and $H^\text{GT}_{\text{f}}$ is therefore,
\begin{equation}\label{eq:IsingPathHam}
 \mathcal{H}^\text{GT}(\lambda) = (1-\lambda) H^\text{GT}_{\text{i}} + \lambda H^\text{GT}_{\text{f}},
\end{equation}
where we choose the path
\begin{equation}\label{eq:smooth_path}
 \lambda(t) = \sin^2 \left( \frac{\pi t}{2 T} \right),
\end{equation}
such that the rate of change of the path $\dot{\lambda}(t)$ at its beginning and end are zero, so as to smoothly switch on/off the gauge potential at these points.

\begin{figure}
\centering
\includegraphics[width=0.95\linewidth]{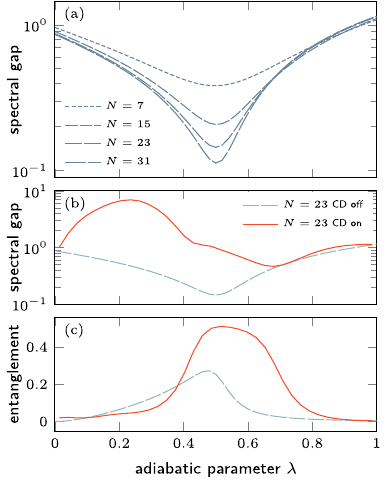}
\caption{\label{fig:5}
(a) The spectral gap of $\mathcal{H}^\text{GT}(\lambda)$ defined in Eq.~\eqref{eq:IsingPathHam} as a function of the adiabatic parameter $\lambda$ for a range of qubit counts $N \in \{7, 15, 23, 31\}$.
(b) The spectral gap and (c) ground-state entanglement entropy for the $N = 23$ qubit Hamiltonian with and without counterdiabatic driving (CD).
In this example $\tilde{A}^{(\chi)}(\lambda)$ is calculated using the method presented in Sec.~\ref{subsec:Methods:1} with a bond dimension $\chi = 4$.
The entanglement entropy corresponds to that of the matrix-product-state ground state found using standard density-matrix renormalization group (DMRG) algorithms~\cite{Sc05}.
}
\end{figure}

The energy difference between the ground and first excited state of $\mathcal{H}^\text{GT}(\lambda)$, i.e.\ the spectral gap, is plotted as a function of $\lambda$ for system sizes $N \in \{7, 15, 23, 31\}$ in Fig.~\ref{fig:5}~(a).
We observe that the minimum spectral gap $\Delta$ decreases as the number of qubits increases. 

To proceed, we calculate the AGPs for system sizes $N \in \{7, 15, 23, 31\}$, variational MPO bond dimensions $\chi \in \{4, 8\}$ and total evolution times $T \in \{0.1, 0.2, 0.3\}$ using the method of Sec.~\ref{subsec:Methods:1}.
More precisely, we discretize the total time $T$ into $S T$ slices where, in all cases $S = 120$, and calculate the AGPs $\tilde{A}^{(\chi)}_{s}$ for each slice.

Since $\dot{\lambda}(t=0) = \dot{\lambda}(t=T) = 0$, the Hamiltonian with counterdiabatic driving $H^\text{GT}(\lambda) = \mathcal{H}^\text{GT}(\lambda) + \dot{\lambda} \tilde{A}^\text{GT}(\lambda)$ still starts with $H^\text{GT}_{\text{i}}$ and ends at $H^\text{GT}_{\text{f}}$ but differs from $\mathcal{H}^\text{GT}(\lambda)$ for $0 < \lambda < 1$.
The spectral gap of $H^\text{GT}(\lambda)$ as a function of $\lambda$ for $N = 23$, $\chi = 4$ and $T = 0.3$ is illustrated in Fig.~\ref{fig:5}~(b), where we observe that the effect of introducing the gauge potential is to increase the spectral gap as compared to $\mathcal{H}^\text{GT}(\lambda)$ alone.
Additionally we observe in Fig.~\ref{fig:5}~(c) that the maximum entanglement entropy of the ground state of $H^\text{GT}(\lambda)$ along the adiabatic path is increased compared to that of $\mathcal{H}^\text{GT}(\lambda)$.
Here, the measure of entanglement used is the von-Neumann entropy $S_\text{VN} = -\sum_{j} \xi_j \ln \xi_j$ where $\xi_j$ are the Schmidt coefficients resulting from the bipartition of the matrix-product-state ground state at the center of the chain.

Given the choice of interpolating Hamiltonian, the initial eigenstate, i.e.\ the ground state of $H^\text{GT}(\lambda=0) = H^\text{GT}_{\text{i}}$, corresponds to the product state $\ket{\psi_{\text{i}}} = \ket{-}^{\otimes N}$ and can easily be incorporated into the cost function $C_{\ket{\psi_{\text{i}}}}(\boldsymbol{\theta})$ for the first chunk of the circuit optimization procedure.
We optimize brickwork circuits of $M = 2$ chunks and $L = 4$ layers using the method of Sec.~\ref{subsec:Methods:2}.
After completing all circuit optimizations we are left with a pair of circuits $U^{(4)}_{m}$ for $m \in \{1, 2\}$ for each parameter set $\{T, \chi, N\}$.

To compare the performance of the counterdiabatic, classically optimized circuits to more traditional techniques we also construct circuits using a second-order Trotter product formula~\cite{HaSu05} of the adiabatically interpolating Hamiltonian defined by $\mathcal{H}^\text{GT}(\lambda)$, i.e., the Trotter circuits do not implement counterdiabatic driving.
Importantly, we use $R = 8$ Trotter layers such that these circuits contain the same number of two-qubit $U_{zz}(\theta)$ gates as the classically optimized counterdiabatic circuits of $R = M \times L = 8$ layers.
To approximate the adiabatic time dynamics as a Trotter circuit $U^\text{Trotter}$, we divide the total adiabatic evolution time $T$ into $R = 8$ equal time chunks labelled by $r = 1, 2, \dots, R$.
We then construct a second-order Trotter approximation of $U^\text{Trotter}_r \approx e^{-\text{i} T \mathcal{H}^\text{GT}(\lambda(t_{r})) / R}$ keeping $\mathcal{H}^\text{GT}$ fixed during each chunk, where $t_{r}$ is the time corresponding the the midpoint of the $r^\text{th}$ chunk.
The full Trotterized adiabatic evolution is then given by the time-ordered product $U^\text{Trotter} = \prod_{r = 1}^{8} U^\text{Trotter}_{r}$.

\begin{figure*}
\centering
\includegraphics[width=0.95\linewidth]{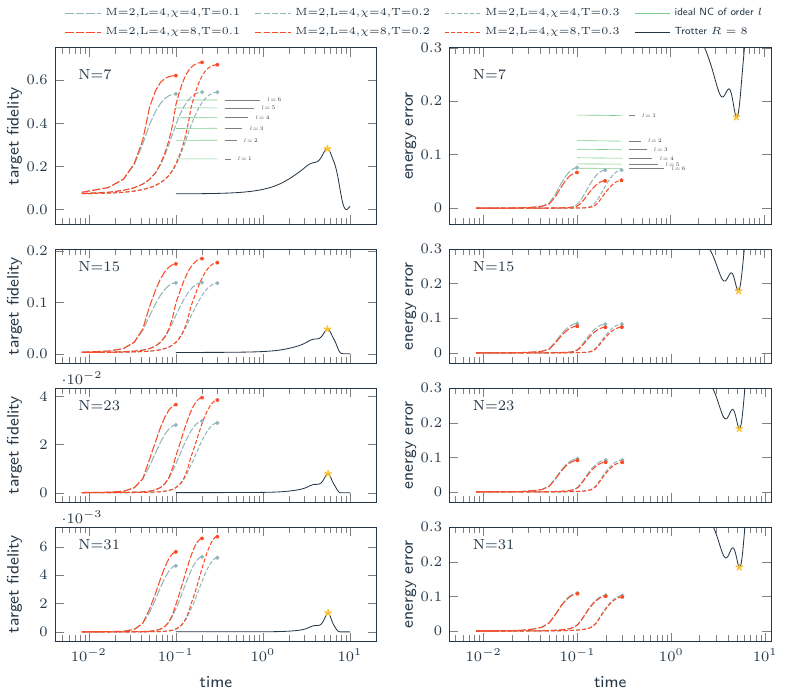}
\caption{\label{fig:6}
Target fidelity~\eqref{eq:TargetFidelity} (left column) and energy errors~\eqref{eq:TargEnergyError} and~\eqref{eq:InstEnergyError} (right column) achieved using either a second-order Trotterization of the standard adiabatic path or classically optimized counterdiabatic circuits for the quantum Ising gap traversal problem defined in Sec.~\ref{subsec:Results:2}.
For each system size $N$, all circuits considered contain the same number of two-qubit gates $R (N-1)$ where $R = M \times L = 8$.
The solid black line indicates the target fidelity and energy error achieved by the Trotter circuit at the \emph{end} of a standard adiabatic sweep, i.e.\ at $\lambda(T) = 1$, for total evolution times $T$ up to a maximum $T = 10$.
The best values achieved by the Trotter circuits are indicated by a star symbol.
The dashed blue and red lines indicate the target fidelities or instantaneous energy errors achieved during the adiabatic evolution using the classically optimized counterdiabatic circuits where the counterdiabatic gauge potentials are calculated as MPOs of bond dimension $\chi$.
The maximum target fidelities and minimum energy errors are achieved at the end of the adiabatic path and are indicated by blue circles (red pentagons) for $\chi = 4$ ($\chi = 8$).
Additionally, the solid green lines indicate the target fidelities and energy errors achieved by ideal adiabatic evolution aided by an order $l$ nested commutator (NC) gauge potential.
}
\end{figure*}

In order to quantify the errors due to imperfect adiabatic evolution, we compute the fidelity between the state prepared by each circuit and the corresponding exact Hamiltonian ground states.
We also compute the relative error in the energies of these states.
These values are defined in terms of the exact ground state of the Hamiltonian $H_s$ at the end of slice $s$, denoted $\ket{\psi_{s}}$, as well as those of the initial and final Hamiltonians, denoted $\ket{\psi_{\text{i}}}$ and $\ket{\psi_{\text{f}}}$ respectively.
We calculate these ground states using standard, numerically exact, density-matrix renormalization group (DMRG) methods.
After applying the adiabatic evolution circuits to the initial state, we are left with $\ket{\phi_{s}} = U_{s}\ket{\psi_{\text{i}}}$ where $U_{s}$ is the circuit preparing the state at the end of slice $s$.
The target fidelity is defined as
\begin{equation}\label{eq:TargetFidelity}
 \mathcal{F}^\text{targ}_{s} = \lvert\bra{\phi_{s}}\ket{\psi_{\text{f}}}\rvert^2
\end{equation}
and the target energy error is given by the relative error
\begin{equation}\label{eq:TargEnergyError}
 \mathcal{E}^\text{targ}_{s} = \frac{\bra{\phi_{s}}H_{\text{f}}\ket{\phi_{s}} - \bra{\psi_{\text{f}}}H_{\text{f}}\ket{\psi_{\text{f}}}}{\bra{\psi_{\text{f}}}H_{\text{f}}\ket{\psi_{\text{f}}}}.
\end{equation}
We also consider the instantaneous energy error defined by
\begin{equation}\label{eq:InstEnergyError}
 \mathcal{E}^\text{inst}_{s} = \frac{\bra{\phi_{s}}H_{s}\ket{\phi_{s}} - \bra{\psi_{s}}H_{s}\ket{\psi_{s}}}{\bra{\psi_{s}}H_{s}\ket{\psi_{s}}},
\end{equation}
where we note that $\mathcal{E}^\text{targ}_{s}$ and $\mathcal{E}^\text{inst}_{s}$ are equivalent at the final time slice. 

The results are presented in Fig.~\ref{fig:6}.
For the parameters chosen, we observe that classically optimized counterdiabatic circuits outperform the non-counterdiabatic Trotter circuits.
In terms of both the target fidelity and energy error, the value obtained is improved using a larger bond dimension for the variationally optimized MPO gauge potential.
Furthermore, for $N = 7$ qubits we observe that classically optimized variational circuits outperform the ideal NC dynamics for all orders $l \leq 6$.
Across all system sizes we observe that classically optimized counterdiabatic circuits outperform the best Trotter result by up to a factor of $5$ for the target fidelities and up to a factor of $3$ for the ground state energy errors.
Note that the energy error at the end of the adiabatic protocol in Fig.~\ref{fig:6} appears to be nearly independent of the system size $N$ and, motivated by the similarity with previous results~\cite{LuEtAl11}, we conjecture that this observation is due to the geometric locality of the terms in the Hamiltonian~\eqref{eq:IsingHamiltonian} whose expectation values give the energy error.

\subsection{Quantum circuits for adiabatic ground state preparation}
\label{subsec:Results:3}

In this section our goal is to compare the capability of our algorithms to adiabatically prepare ground states with the popular tensor network based approach of~\cite{Ra20}.
To achieve high state preparation fidelities, here we utilize that the final state accuracy after adiabatic evolution can be systematically improved simply by increasing the total evolution time $T$.
For large $T$ the rate of change of the adiabatic parameter $\dot{\lambda}$ is small such that the counterdiabatic driving is weak and can be ignored.

In the following, we investigate the ability of our methods to find circuits which follow the instantaneous ground state along an adiabatic path defined by the initial Hamiltonian $H^{\text{C}}_{\text{i}} = H_{\text{Ising}}(0,-1,0)$ and the final Hamiltonian $H^{\text{C}}_{\text{f}}=H_{\text{Ising}}(1,-1,0)$ such that
\begin{equation}\label{eq:InterpHamCrit}
 \mathcal{H}^{\text{C}}(\lambda) = (1-\lambda)H_{\text{i}}^{C}+\lambda H_{\text{f}}^C,
\end{equation}
where we choose the path
\begin{equation}
 \lambda(t) = \sin^2\left(\frac{\pi}{2}\sin^2\left(\frac{\pi t}{2T}\right)\right).
\end{equation}
The ground state of the transverse-field Ising Hamiltonian $H^{C}_{\text{f}}$ corresponds to the critical point in the thermodynamic limit.

\begin{figure}
\centering
\includegraphics[width=0.95\linewidth]{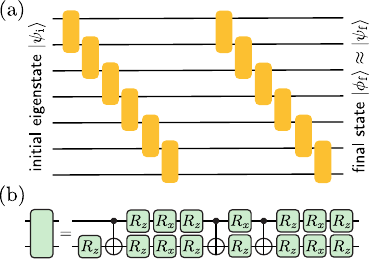}
\caption{\label{fig:7}
(a) A sequential circuit ansatz for $N = 7$ qubits and $L = 2$ layers.
(b) Substructure of the two-qubit blocks of the sequential circuit where three CNOT gates and fifteen one-qubit rotations parameterize an arbitrary two-qubit unitary operator~\cite{ShMaBu04}.
}
\end{figure}

As an ansatz we choose the sequential circuit structure shown graphically in Fig.~\ref{fig:7}~(a) where the two-qubit blocks have the substructure shown in Fig.~\ref{fig:7}~(b).
At each time slice $s = 1, 2, \dots, ST$, the short-time evolution operator is approximated by a second-order Trotter product formula, which, for the interpolating Hamiltonian~\eqref{eq:InterpHamCrit}, can be represented efficiently as an MPO with bond dimension two.

We consider an $N = 24$ qubit system and optimize the circuit using a single ($M = 1$) chunk, $L = 2$ layers of the sequential ansatz and a total adiabatic time $T = 64$ with $S = 8$ slices per unit time, i.e.\ a Trotter time step $\tau = 0.125$.
Additionally, we examine the convergence properties of our optimization routine by varying the number of L-BFGS iterations per slice $Q \in \{10, 50, 100\}$.

\begin{figure}
\centering
\includegraphics[width=0.95\linewidth]{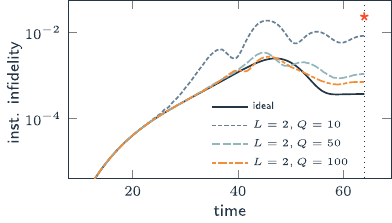}
\caption{\label{fig:8}
Instantaneous infidelity~\eqref{eq:InstInfidelity} as a function of time for $L = 2$ layer classically optimized circuits acting on $N = 24$ qubits, alongside the ideal adiabatic Trotter evolution.
The total adiabatic time is $T = 64$ and there are $S = 8$ slices per unit time.
The star symbol indicates the target infidelity achieved using a nine-layer sequential circuit by the method of~\cite{Ra20}.
}
\end{figure}

Plotted in Fig.~\ref{fig:8} is the instantaneous infidelity
\begin{equation}\label{eq:InstInfidelity}
 \mathcal{I}^\text{inst}_{s} = 1 - \lvert\bra{\phi_{s}}\ket{\psi_{s}}\rvert^2,
\end{equation}
where $\ket{\phi_{s}}$ is the state prepared by the classically optimized circuit (or the ideal adiabatic evolution explained below), and $\ket{\psi_{s}}$ is the instantaneous ground state calculated using numerically exact DMRG methods.
In addition to the instantaneous infidelities associated with the classically optimized circuits, we also plot that of the ideal adiabatic evolution calculated by evolving the initial state along the adiabatic path using the same Trotter circuit in a numerically exact way.
We observe that the classically optimized circuit of $L = 2$ layers maps the initial state to one which closely follows the ideal evolution and the degree to which it does so is dependent on the number of L-BFGS iterations $Q$ performed in each slice.
Using just $L = 2$ sequential circuit layers, our approach achieves a target state infidelity which is more than one order of magnitude lower than that achieved using a nine-layer sequential circuit by the method of~\cite{Ra20}, which is indicated by the star symbol at time $T = 64$ in Fig.~\ref{fig:8}.

\section{Discussion}
\label{sec:Discussion}

In this article, we numerically demonstrate that adiabatic evolution with counterdiabatic driving for specific, one-dimensional quantum many-body systems can be accurately represented by appending shallow one-dimensional quantum circuits that are optimized using standard tensor network techniques.
We note that the specific quantum many-body ground states considered throughout this article can alternatively be calculated on classical computers with high accuracies using DMRG and the resulting matrix product states can directly be transformed into quantum circuits~\cite{ScEtAl05, ScEtAl07, LuEtAl20, Ra20, AkEtAl23, MaStWeCi24}.
We have provided one example in Sec.~\ref{subsec:Results:3} that shows how our methods can outperform this approach and therefore constitute a valuable addition to the arsenal of tensor network techniques available for this purpose.
We anticipate that our adiabatic quantum computing procedure will outperform fully classical approaches in the preparation of highly entangled states for which MPS fail, e.g.\ large two-dimensional quantum many-body systems~\cite{StWh12} or quantum circuits representing adiabatic evolution for combinatorial optimization~\cite{DuEtAl22a, DuEtAl22b}.

Our approach can be readily applied to two-dimensional quantum many-body systems by making use of the corresponding two-dimensional tensor-network algorithms~\cite{VeMuCi08, Or14, Ba23}.
More specifically, shallow two-dimensional quantum circuits can be expressed in terms of so-called projected entangled pair states~\cite{VeCi04} for which powerful optimization methods are known, e.g.~\cite{VeCi04, LuCiBa14a, LuCiBa14b, CzDzCo19, McSz21}.

\begin{figure}
\centering
\includegraphics[width=0.95\linewidth]{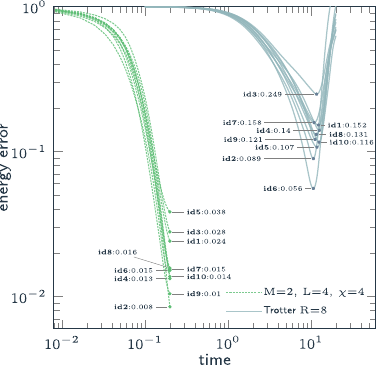}
\caption{\label{fig:9}
Energy error achieved by circuits of $R = M \times L = 8$ layers for random instances, $\textbf{\textrm{id}} \in \{1, 2, \dots, 10\}$, of the nearest-neighbour Ising Hamiltonian~\eqref{eq:IsingHamiltonian} with $N = 8$ qubits. 
For the initial Hamiltonian we choose $g_k = 0.5\; \forall k$ and $J_k = h_k = 0\; \forall k$.
For the final Hamiltonian, couplings $J_{k}$ and local fields $h_{k}$ are chosen randomly from the sets $J_{k} \in \pm\{0.2, 0.4, 0.6, 0.8, 1.0\}$, $h_{k} \in \pm\{0.0, 0.2, 0.4, 0.6, 0.8\}$ and $g_k = 0\; \forall k$.
For each of the $10$ instances, the target energy error~\eqref{eq:TargEnergyError} of the classically optimized counterdiabatic circuits along the linear adiabatic path, $\lambda(t) = t/T$ with $T = 0.2$, from $\lambda = 0$ to $\lambda = 1$ is indicated by the dashed green lines.
Additionally, for each instance we plot the energy error achieved at the \emph{end} of the adiabatic path, i.e.\ at $\lambda = 1$ only, by a second-order Trotter circuit following the non-counterdiabatic linear adiabatic path for a range of total times $T$, up to a maximum total time $T = 20$.
In all cases, the minimum energy achieved is indicated by a pentagon (circle) for Trotter (classically optimized counterdiabatic) and is labelled with its instance number alongside the numerical value of the error.
}
\end{figure}

The quantum circuit optimization procedures can also be used to adiabatically solve combinatorial optimization problems, which have historically always been an important application for adiabatic quantum computing~\cite{KaNi98, FaEtAl00, AlLi18, HaEtAl20}.
For one-dimensional classical systems, it makes sense to use the same approach as for the quantum systems and optimize shallow one-dimensional quantum circuits to realize the adiabatic protocol.
Figure~\ref{fig:9} shows results for $10$ random instances of classical nearest-neighbour Ising chains.
We see that, similar to the quantum counterpart, classically optimized circuits with counterdiabatic driving outperform non-counterdiabatic Trotter circuits with the same number of two-qubit gates.
We emphasize, however, that one-dimensional classical spin systems can be efficiently solved using classical computers.

\begin{figure}
\centering
\includegraphics[width=71.645mm]{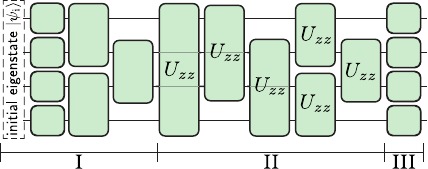}
\caption{\label{fig:10}
Part I is a brickwork circuit of $L = 1$ layer.
The component inside the dashed box needs to be included only in the first chunk optimization of Fig.~\ref{fig:1}.
Part II consists of commuting two-qubit gates acting on all possible pairs of qubits and part III contains generic single-qubit gates.
The gate parameterization is defined in Fig.~\ref{fig:2}.
}
\end{figure}

For the more interesting, hard classical problems with nonlocal interactions, including quadratic unconstrained binary optimization~\cite{Lu14, GlEtAl22}, we propose to use a variational circuit architecture consisting of three parts.
The first part is a circuit structure that can be dealt with efficiently using classical techniques, for instance a shallow brickwork circuit or a tensor-network-inspired quantum circuit as considered, e.g., in~\cite{HaEtAl22, CePlLu23, HaEtAl23}.
The second part is composed of commuting many-qubit gates, e.g.\ one layer of the quantum approximate optimization algorithm (QAOA)~\cite{FaGoGu14} where, in contrast to~\cite{FaGoGu14}, each gate contributes an independent variational parameter as in multi-angle QAOA~\cite{HeEtAl22}.
The third part consists of one layer of arbitrary single-qubit gates.
A quantum circuit consisting of these three parts can be efficiently optimized using classical tensor network algorithms for so-called weighted graph states~\cite{AnEtAl06, AnBrDu07, HaEtAl07, HuEtAl09, HuEtAl11}.
We present an example circuit architecture of this type in Fig.~\ref{fig:10}.
For the classical optimization of such circuit structures to be efficient, however, it needs to be based on expectation values of few-qubit operators~\cite{AnEtAl06, AnBrDu07, HaEtAl07, HuEtAl09, HuEtAl11}.
Therefore, in the context of these circuit architectures, we propose to replace the cost function~\eqref{eq:CostSlice} by the corresponding one derived from McLachlan's variational principle~\cite{Mc64}, see Eqs.~(11) and~(12) in~\cite{YuEtAl19}.
Additionally, to represent the AGP, we propose to replace the variational MPO by a variational few-body ansatz whose nonlocality can be systematically increased~\cite{SePo17, ClEtAl19, HeChSo22, CeEtAl23}.
An interesting possibility is to go beyond previous proposals and study a variational ansatz defined in terms of generic tensors similar to the one in~\cite{LuEtAl16} which has proven to be a successful approach in a different context.

Another exciting future research direction is to explore alternative approaches for the optimization, e.g.\ approaches that aim at preserving local information~\cite{LeEtAl17, WhEtAl18, SuPiTa19, KvHeBa22, FrTaBa24, ArEtAl24}.

\emph{Note added.} A few days before our article appeared on the arXiv, a related article was published there~\cite{KiFiSe23} (see final journal version~\cite{KiFiSe24}) in which the authors also propose to use variational MPOs for AGPs.
While their numerical procedures for the MPO approximation of the AGP are similar to ours, they use them in a different context, namely to improve classical tensor network optimization and scan phase diagrams.
In contrast, in our article we apply the variational MPO approximation of the AGP to the quantum circuit optimization for adiabatic quantum computing.
In this way, the two articles complement each other and we recommend reading both of them.

\appendix

\section{Adiabatic gauge potential}
\label{app:A}

In this Appendix, we provide a brief derivation of the AGP and some of its properties; see~\cite{SePo17, KoEtAl17} for a more comprehensive presentation.

We consider a general Hamiltonian $\mathcal{H}(\lambda)$ depending on a scalar $\lambda$, which can, e.g., be of the particular form $\mathcal{H}(\lambda) = (1-\lambda) H_{\text{i}} + \lambda H_{\text{f}}$ used in Fig.~\ref{fig:1}~(a).
We define the unitary $U(\lambda)$ that diagonalizes $\mathcal{H}(\lambda)$,
\begin{equation}\label{eq:HTilde}
 U^{\dag}(\lambda) \mathcal{H}(\lambda) U(\lambda) = \bar{\mathcal{H}}(\lambda),
\end{equation}
where $\bar{\mathcal{H}}(\lambda)$ is diagonal and contains the eigenvalues of $\mathcal{H}(\lambda)$.
In other words, the $v$th column of $U(\lambda)$ contains the eigenvector $|v(\lambda)\rangle$ of $\mathcal{H}(\lambda)$ to eigenvalue $E_{v}(\lambda)$, which is the $v$th entry on the diagonal of $\bar{\mathcal{H}}(\lambda)$.
This is equivalent to
\begin{eqnarray}\label{eq:H}
 \mathcal{H}(\lambda) & = & U(\lambda) \bar{\mathcal{H}}(\lambda) U^{\dag}(\lambda)\\\label{eq:E}
 & = & \sum_{v} E_{v}(\lambda) |v(\lambda)\rangle \langle v(\lambda)|
\end{eqnarray}
where $(\cdot)^{\dag}$ is the adjoint of $(\cdot)$.
In the adiabatic protocol $\lambda = \lambda(t)$ and so we refer to the eigenstates $|v(\lambda)\rangle$ as the instantaneous eigenstates.
For an arbitrary operator $O$ we define $\bar{O} = U^{\dag}(\lambda) O U(\lambda)$ corresponding to its representation in the instantaneous eigenbasis.
Similarly for an arbitrary state $|\psi\rangle$ we define $|\bar{\psi}\rangle = U^{\dag}(\lambda) |\psi\rangle$.

We want to write the time-dependent Schr\"{o}dinger equation in the instantaneous eigenbasis.
To that end, we consider the time derivative of the components of $|\psi\rangle$ in the instantanous eigenbasis, i.e.\ $\partial_{t} |\bar{\psi}\rangle$, for which we obtain
\begin{eqnarray}
 \text{i} \partial_{t} |\bar{\psi}\rangle & = & \text{i} \partial_{t} \left( U^{\dag} |\psi\rangle \right)\\
 & = & \text{i} \left( \partial_{t} U^{\dag} \right) |\psi\rangle + \text{i} U^{\dag} \left( \partial_{t} |\psi\rangle \right)\\
 & = & \text{i} \left( \partial_{t} \lambda \right) \left( \partial_{\lambda} U^{\dag} \right) U U^{\dag} |\psi\rangle + U^{\dag} \mathcal{H} |\psi\rangle\\
 & = & \text{i} \dot{\lambda} \left( \partial_{\lambda} U^{\dag} \right) U |\bar{\psi}\rangle + U^{\dag} \mathcal{H} U U^{\dag} |\psi\rangle\\
 & = & -\dot{\lambda} \bar{A} |\bar{\psi}\rangle + \bar{\mathcal{H}} |\bar{\psi}\rangle.\label{eq:HGPot}
\end{eqnarray}
Here we have used $\text{i} \partial_{t} |\psi\rangle = \mathcal{H} |\psi\rangle$.
We have defined $\dot{\lambda} = \partial_{t} \lambda$ and the gauge potential
\begin{equation}\label{eq:GPotInst}
 \bar{A} = -\text{i} \left( \partial_{\lambda} U^{\dag} \right) U
\end{equation}
in the instantaneous eigenbasis, which reads
\begin{equation}\label{eq:GPotSt}
 A = -\text{i} U \left( \partial_{\lambda} U^{\dag} \right)
\end{equation}
in the standard eigenbasis.

Equation~\eqref{eq:HGPot} implies that, in the instantaneous eigenbasis the evolution happens according to $\bar{\mathcal{H}} - \dot{\lambda} \bar{A}$.
Therefore, if the initial state is an eigenstate of $\mathcal{H}$, then $\bar{A}$ causes the state to deviate from the instantaneous eigenstate during the evolution.
We counteract this deviation by using $\mathcal{H} + \dot{\lambda} A$ instead of just $\mathcal{H}$ in the Schr\"{o}dinger equation.
Then, assuming the exact $A$ is used, an initial eigenstate of $\mathcal{H}$ propagates as an exact instantaneous eigenstate during the entire evolution.

The AGP is Hermitian,
\begin{eqnarray}
 A^{\dag} & = & \left( -\text{i} U \left( \partial_{\lambda} U^{\dag} \right) \right)^{\dag}\\
 & = & \text{i} \left( \partial_{\lambda} U^{\dag} \right)^{\dag} U^{\dag}\\\label{eq:GPotSt2}
 & = & \text{i} \left( \partial_{\lambda} U \right) U^{\dag}\\
 & = & \text{i} \partial_{\lambda} \left( U U^{\dag} \right) - \text{i} U \left( \partial_{\lambda} U^{\dag} \right)\\
 & = & -\text{i} U \left( \partial_{\lambda} U^{\dag} \right)\\
 & = & A,
\end{eqnarray}
where we have used $\partial_{\lambda}(U U^{\dag}) = \partial_{\lambda} \mathds{1} = 0$ and $\mathds{1}$ is the identity operator.
Note that, for a real-valued Hamiltonian $\mathcal{H}$, $U$ can be chosen to be a real-valued orthogonal matrix and then $A$ contains only imaginary numbers.

In terms of the instantaneous eigenstates $| v \rangle$ and eigenenergies $E_{v}$, $A$ in Eq.~\eqref{eq:GPotSt} takes on the form
\begin{eqnarray}
 A & = & -\text{i} \sum_{v} | v \rangle \left( \partial_{\lambda} \langle v | \right)\\
 & = & -\text{i} \sum_{v} \sum_{w \neq v} \frac{\langle v | \left( \partial_{\lambda} \mathcal{H} \right) | w \rangle}{E_{v} - E_{w}} | v \rangle \langle w |,
\end{eqnarray}
where we have used $( \partial_{\lambda} \mathcal{H} )^{\dag} = \partial_{\lambda} \mathcal{H}$ and
\begin{equation}
 \partial_{\lambda} | v \rangle = \sum_{w \neq v} \frac{\langle w | \left( \partial_{\lambda} \mathcal{H} \right) | v \rangle}{E_{v} - E_{w}} | w \rangle.
\end{equation}
This follows, e.g., from differentiating the time-independent Schr\"{o}dinger equation $\mathcal{H} | v \rangle = E_{v} | v \rangle$ with respect to $\lambda$,
\begin{equation}
 \left( \partial_{\lambda} \mathcal{H} \right) | v \rangle + \mathcal{H} \left( \partial_{\lambda} | v \rangle \right) = \left( \partial_{\lambda} E_{v} \right) | v \rangle + E_{v} \left( \partial_{\lambda} | v \rangle \right),
\end{equation}
which, multiplied from the left by $\langle w |$, leads to
\begin{equation}
 \langle w | \left( \partial_{\lambda} | v \rangle \right) = \frac{\langle w | \left( \partial_{\lambda} \mathcal{H} \right) | v \rangle}{E_{v} - E_{w}}
\end{equation}
for all $w \neq v$.
For $w = v$, we remember that $A$ is Hermitian and for a real-valued Hamiltonian $\mathcal{H}$ we can choose $A$ to contain only imaginary numbers, so that
\begin{eqnarray}
 \langle v | A | v \rangle & = & \langle v | A^{\dag} | v \rangle\\
 & = & \left( \langle v | A | v \rangle \right)^{*}\\
 & = & 0
\end{eqnarray}
where $(\cdot)^{*}$ is the complex conjugate of $(\cdot)$.

\section{Variational gauge potential}
\label{app:B}

In this Appendix, we present a concise derivation of the cost function that we minimize in the main text using a variational ansatz for the AGP; see~\cite{SePo17, KoEtAl17} for further details.

We define the operators
\begin{eqnarray}\label{eq:G}
 G & = & \left( \partial_{\lambda} \mathcal{H} \right) + \text{i} [A, \mathcal{H}],\\
 M & = & \sum_{v} \left( \partial_{\lambda} E_{v} \right) | v \rangle \langle v |,
\end{eqnarray}
where $[O_{1}, O_{2}]$ is the commutator of $O_{1}$ and $O_{2}$.
The operators $G$ and $M$ are equivalent to each other, i.e.\ $G = M$, for the exact gauge potential $A$.
This can be seen by differentiating $\mathcal{H}$ with respect to $\lambda$,
\begin{eqnarray}
 \partial_{\lambda} \mathcal{H} & = & \partial_{\lambda} \left( U \bar{\mathcal{H}} U^{\dag} \right)\\\nonumber
 & = & \left( \partial_{\lambda} U \right) \bar{\mathcal{H}} U^{\dag} + U \left( \partial_{\lambda} \bar{\mathcal{H}} \right) U^{\dag} +\\
 & & U \bar{\mathcal{H}} \left( \partial_{\lambda} U^{\dag} \right)\\\nonumber
 & = & \left( \partial_{\lambda} U \right) U^{\dag} \mathcal{H} + \sum_{v} \left( \partial_{\lambda} E_{v} \right) | v \rangle \langle v | +\\
 & & \mathcal{H} U \left( \partial_{\lambda} U^{\dag} \right)\\
 & = & -\text{i} A \mathcal{H} + \sum_{v} \left( \partial_{\lambda} E_{v} \right) | v \rangle \langle v | + \text{i} \mathcal{H} A\\\label{eq:HDeriv}
 & = & -\text{i} [A, \mathcal{H}] + \sum_{v} \left( \partial_{\lambda} E_{v} \right) | v \rangle \langle v |,
\end{eqnarray}
where we have used Eqs.~\eqref{eq:HTilde}, \eqref{eq:H}, \eqref{eq:E}, \eqref{eq:GPotSt} and \eqref{eq:GPotSt2}.
Note that, because $M$ commutes with $\mathcal{H}$, $G$ commutes with $\mathcal{H}$ for the exact gauge potential $A$.

Following~\cite{SePo17, KoEtAl17}, we define as the goal of the variational procedure the optimization of a variational ansatz $\tilde{A}$ for $A$ in $G(A)$, see Eq.~\eqref{eq:G}, to minimize the Frobenius distance
\begin{eqnarray}\label{eq:FrobDist}
 \epsilon(\tilde{A}) & = & \text{tr} \left[ \left( G(\tilde{A}) - M \right)^{\dag} \left( G(\tilde{A}) - M \right) \right]\\\nonumber
 & = & \text{tr} \left[ G^{\dag}(\tilde{A}) G(\tilde{A}) \right] - 2 \Re \{ \text{tr}\left[ G^{\dag}(\tilde{A}) M \right] \} +\\
 & & \text{tr} \left[ M^{\dag} M \right]
\end{eqnarray}
where $\text{tr}[\cdot]$ is the trace of $[\cdot]$ and $\Re \{ \cdot \}$ is the real part of $\{ \cdot \}$.
We see that
\begin{eqnarray}
 \text{tr} \left[ G^{\dag} M \right] & = & \text{tr} \left[ \left( \left( \partial_{\lambda} \mathcal{H} \right) + \text{i} [\tilde{A}, \mathcal{H}] \right)^{\dag} M \right]\\\nonumber
 & = & \text{tr} \left[ \left( \partial_{\lambda} \mathcal{H} \right)^{\dag} M \right] - \text{i} \text{tr} \left[ \mathcal{H} \tilde{A}^{\dag} M \right] +\\
 & & \text{i} \text{tr} \left[ \tilde{A}^{\dag} \mathcal{H} M \right]\\\nonumber
 & = & \text{tr} \left[ \left( \partial_{\lambda} \mathcal{H} \right)^{\dag} M \right] - \text{i} \text{tr} \left[ \tilde{A}^{\dag} M \mathcal{H} \right] +\\
 & & \text{i} \text{tr} \left[ \tilde{A}^{\dag} M \mathcal{H} \right]\\\nonumber
 & = & \text{tr} \left[ \left( \partial_{\lambda} \mathcal{H} \right)^{\dag} M \right]\\
 & = & \text{tr} \left[ M^{\dag} M \right],
\end{eqnarray}
where we have used the cyclic property of the trace and that $M$ commutes with $\mathcal{H}$.
Also, we have inserted Eq.~\eqref{eq:HDeriv} and evaluated the resulting expression.
Therefore the Frobenius distance in Eq.~\eqref{eq:FrobDist} is
\begin{eqnarray}
 \epsilon(\tilde{A}) & = & \text{tr}[G^{\dag}(\tilde{A}) G(\tilde{A})] - \text{tr}[M^{\dag} M].
\end{eqnarray}
Because $M$ does not depend on $\tilde{A}$, we minimize the Frobenius distance in Eq.~\eqref{eq:FrobDist} variationally via $\tilde{A}$ by minimizing the cost function
\begin{equation}\label{eq:Cost}
 c(\tilde{A}) = \text{tr} \left[ G^{\dag}(\tilde{A}) G(\tilde{A}) \right].
\end{equation}
This cost function can be alternatively derived starting from the system of linear equations
\begin{equation}\label{eq:LE}
 \left[ \mathcal{H}, \tilde{A} \right] = -\text{i} \left( \partial_{\lambda} \mathcal{H} \right)
\end{equation}
and solving them for a variational $\tilde{A}$ by minimizing the Frobenius distance between the left-hand and right-hand side of Eq.~\eqref{eq:LE}.
Since tensor network algorithms for linear equations exist~\cite{OsDo12, LuMoJa18}, we use Eq.~\eqref{eq:LE} as the starting point for the variational MPO optimization in the main text.

\vspace{1mm}

\bibliography{bibliography}

\begin{thebibliography}{100}%
\makeatletter
\providecommand \@ifxundefined [1]{%
 \@ifx{#1\undefined}
}%
\providecommand \@ifnum [1]{%
 \ifnum #1\expandafter \@firstoftwo
 \else \expandafter \@secondoftwo
 \fi
}%
\providecommand \@ifx [1]{%
 \ifx #1\expandafter \@firstoftwo
 \else \expandafter \@secondoftwo
 \fi
}%
\providecommand \natexlab [1]{#1}%
\providecommand \enquote  [1]{``#1''}%
\providecommand \bibnamefont  [1]{#1}%
\providecommand \bibfnamefont [1]{#1}%
\providecommand \citenamefont [1]{#1}%
\providecommand \href@noop [0]{\@secondoftwo}%
\providecommand \href [0]{\begingroup \@sanitize@url \@href}%
\providecommand \@href[1]{\@@startlink{#1}\@@href}%
\providecommand \@@href[1]{\endgroup#1\@@endlink}%
\providecommand \@sanitize@url [0]{\catcode `\\12\catcode `\$12\catcode
  `\&12\catcode `\#12\catcode `\^12\catcode `\_12\catcode `\%12\relax}%
\providecommand \@@startlink[1]{}%
\providecommand \@@endlink[0]{}%
\providecommand \url  [0]{\begingroup\@sanitize@url \@url }%
\providecommand \@url [1]{\endgroup\@href {#1}{\urlprefix }}%
\providecommand \urlprefix  [0]{URL }%
\providecommand \Eprint [0]{\href }%
\providecommand \doibase [0]{https://doi.org/}%
\providecommand \selectlanguage [0]{\@gobble}%
\providecommand \bibinfo  [0]{\@secondoftwo}%
\providecommand \bibfield  [0]{\@secondoftwo}%
\providecommand \translation [1]{[#1]}%
\providecommand \BibitemOpen [0]{}%
\providecommand \bibitemStop [0]{}%
\providecommand \bibitemNoStop [0]{.\EOS\space}%
\providecommand \EOS [0]{\spacefactor3000\relax}%
\providecommand \BibitemShut  [1]{\csname bibitem#1\endcsname}%
\let\auto@bib@innerbib\@empty
%</preamble>
\bibitem [{\citenamefont {Kadowaki}\ and\ \citenamefont
  {Nishimori}(1998)}]{KaNi98}%
  \BibitemOpen
  \bibfield  {author} {\bibinfo {author} {\bibfnamefont {T.}~\bibnamefont
  {Kadowaki}}\ and\ \bibinfo {author} {\bibfnamefont {H.}~\bibnamefont
  {Nishimori}},\ }\bibfield  {title} {\bibinfo {title} {Quantum annealing in
  the transverse {Ising} model},\ }\href
  {https://doi.org/10.1103/PhysRevE.58.5355} {\bibfield  {journal} {\bibinfo
  {journal} {Phys. Rev. E}\ }\textbf {\bibinfo {volume} {58}},\ \bibinfo
  {pages} {5355} (\bibinfo {year} {1998})}\BibitemShut {NoStop}%
\bibitem [{\citenamefont {Farhi}\ \emph {et~al.}(2000)\citenamefont {Farhi},
  \citenamefont {Goldstone}, \citenamefont {Gutmann},\ and\ \citenamefont
  {Sipser}}]{FaEtAl00}%
  \BibitemOpen
  \bibfield  {author} {\bibinfo {author} {\bibfnamefont {E.}~\bibnamefont
  {Farhi}}, \bibinfo {author} {\bibfnamefont {J.}~\bibnamefont {Goldstone}},
  \bibinfo {author} {\bibfnamefont {S.}~\bibnamefont {Gutmann}},\ and\ \bibinfo
  {author} {\bibfnamefont {M.}~\bibnamefont {Sipser}},\ }\href@noop {}
  {\bibinfo {title} {{Quantum Computation by Adiabatic Evolution}}} (\bibinfo
  {year} {2000}),\ \Eprint {https://arxiv.org/abs/quant-ph/0001106}
  {arXiv:quant-ph/0001106 [quant-ph]} \BibitemShut {NoStop}%
\bibitem [{\citenamefont {Albash}\ and\ \citenamefont {Lidar}(2018)}]{AlLi18}%
  \BibitemOpen
  \bibfield  {author} {\bibinfo {author} {\bibfnamefont {T.}~\bibnamefont
  {Albash}}\ and\ \bibinfo {author} {\bibfnamefont {D.~A.}\ \bibnamefont
  {Lidar}},\ }\bibfield  {title} {\bibinfo {title} {Adiabatic quantum
  computation},\ }\href {https://doi.org/10.1103/RevModPhys.90.015002}
  {\bibfield  {journal} {\bibinfo  {journal} {Rev. Mod. Phys.}\ }\textbf
  {\bibinfo {volume} {90}},\ \bibinfo {pages} {015002} (\bibinfo {year}
  {2018})}\BibitemShut {NoStop}%
\bibitem [{\citenamefont {Hauke}\ \emph {et~al.}(2020)\citenamefont {Hauke},
  \citenamefont {Katzgraber}, \citenamefont {Lechner}, \citenamefont
  {Nishimori},\ and\ \citenamefont {Oliver}}]{HaEtAl20}%
  \BibitemOpen
  \bibfield  {author} {\bibinfo {author} {\bibfnamefont {P.}~\bibnamefont
  {Hauke}}, \bibinfo {author} {\bibfnamefont {H.~G.}\ \bibnamefont
  {Katzgraber}}, \bibinfo {author} {\bibfnamefont {W.}~\bibnamefont {Lechner}},
  \bibinfo {author} {\bibfnamefont {H.}~\bibnamefont {Nishimori}},\ and\
  \bibinfo {author} {\bibfnamefont {W.~D.}\ \bibnamefont {Oliver}},\ }\bibfield
   {title} {\bibinfo {title} {Perspectives of quantum annealing: methods and
  implementations},\ }\href {https://doi.org/10.1088/1361-6633/ab85b8}
  {\bibfield  {journal} {\bibinfo  {journal} {Rep. Prog. Phys.}\ }\textbf
  {\bibinfo {volume} {83}},\ \bibinfo {pages} {054401} (\bibinfo {year}
  {2020})}\BibitemShut {NoStop}%
\bibitem [{\citenamefont {Johnson}\ \emph {et~al.}(2011)\citenamefont
  {Johnson}, \citenamefont {Amin}, \citenamefont {Gildert}, \citenamefont
  {Lanting}, \citenamefont {Hamze}, \citenamefont {Dickson}, \citenamefont
  {Harris}, \citenamefont {Berkley}, \citenamefont {Johansson}, \citenamefont
  {Bunyk}, \citenamefont {Chapple}, \citenamefont {Enderud}, \citenamefont
  {Hilton}, \citenamefont {Karimi}, \citenamefont {Ladizinsky}, \citenamefont
  {Ladizinsky}, \citenamefont {Oh}, \citenamefont {Perminov}, \citenamefont
  {Rich}, \citenamefont {Thom}, \citenamefont {Tolkacheva}, \citenamefont
  {Truncik}, \citenamefont {Uchaikin}, \citenamefont {Wang}, \citenamefont
  {Wilson},\ and\ \citenamefont {Rose}}]{JoEtAl11}%
  \BibitemOpen
  \bibfield  {author} {\bibinfo {author} {\bibfnamefont {M.~W.}\ \bibnamefont
  {Johnson}}, \bibinfo {author} {\bibfnamefont {M.~H.~S.}\ \bibnamefont
  {Amin}}, \bibinfo {author} {\bibfnamefont {S.}~\bibnamefont {Gildert}},
  \bibinfo {author} {\bibfnamefont {T.}~\bibnamefont {Lanting}}, \bibinfo
  {author} {\bibfnamefont {F.}~\bibnamefont {Hamze}}, \bibinfo {author}
  {\bibfnamefont {N.}~\bibnamefont {Dickson}}, \bibinfo {author} {\bibfnamefont
  {R.}~\bibnamefont {Harris}}, \bibinfo {author} {\bibfnamefont {A.~J.}\
  \bibnamefont {Berkley}}, \bibinfo {author} {\bibfnamefont {J.}~\bibnamefont
  {Johansson}}, \bibinfo {author} {\bibfnamefont {P.}~\bibnamefont {Bunyk}},
  \bibinfo {author} {\bibfnamefont {E.~M.}\ \bibnamefont {Chapple}}, \bibinfo
  {author} {\bibfnamefont {C.}~\bibnamefont {Enderud}}, \bibinfo {author}
  {\bibfnamefont {J.~P.}\ \bibnamefont {Hilton}}, \bibinfo {author}
  {\bibfnamefont {K.}~\bibnamefont {Karimi}}, \bibinfo {author} {\bibfnamefont
  {E.}~\bibnamefont {Ladizinsky}}, \bibinfo {author} {\bibfnamefont
  {N.}~\bibnamefont {Ladizinsky}}, \bibinfo {author} {\bibfnamefont
  {T.}~\bibnamefont {Oh}}, \bibinfo {author} {\bibfnamefont {I.}~\bibnamefont
  {Perminov}}, \bibinfo {author} {\bibfnamefont {C.}~\bibnamefont {Rich}},
  \bibinfo {author} {\bibfnamefont {M.~C.}\ \bibnamefont {Thom}}, \bibinfo
  {author} {\bibfnamefont {E.}~\bibnamefont {Tolkacheva}}, \bibinfo {author}
  {\bibfnamefont {C.~J.~S.}\ \bibnamefont {Truncik}}, \bibinfo {author}
  {\bibfnamefont {S.}~\bibnamefont {Uchaikin}}, \bibinfo {author}
  {\bibfnamefont {J.}~\bibnamefont {Wang}}, \bibinfo {author} {\bibfnamefont
  {B.}~\bibnamefont {Wilson}},\ and\ \bibinfo {author} {\bibfnamefont
  {G.}~\bibnamefont {Rose}},\ }\bibfield  {title} {\bibinfo {title} {Quantum
  annealing with manufactured spins},\ }\href
  {https://doi.org/10.1038/nature10012} {\bibfield  {journal} {\bibinfo
  {journal} {Nature}\ }\textbf {\bibinfo {volume} {473}},\ \bibinfo {pages}
  {194} (\bibinfo {year} {2011})}\BibitemShut {NoStop}%
\bibitem [{\citenamefont {Farhi}\ \emph {et~al.}(2014)\citenamefont {Farhi},
  \citenamefont {Goldstone},\ and\ \citenamefont {Gutmann}}]{FaGoGu14}%
  \BibitemOpen
  \bibfield  {author} {\bibinfo {author} {\bibfnamefont {E.}~\bibnamefont
  {Farhi}}, \bibinfo {author} {\bibfnamefont {J.}~\bibnamefont {Goldstone}},\
  and\ \bibinfo {author} {\bibfnamefont {S.}~\bibnamefont {Gutmann}},\
  }\href@noop {} {\bibinfo {title} {{A Quantum Approximate Optimization
  Algorithm}}} (\bibinfo {year} {2014}),\ \Eprint
  {https://arxiv.org/abs/1411.4028} {arXiv:1411.4028 [quant-ph]} \BibitemShut
  {NoStop}%
\bibitem [{\citenamefont {Zener}(1932)}]{Ze32}%
  \BibitemOpen
  \bibfield  {author} {\bibinfo {author} {\bibfnamefont {C.}~\bibnamefont
  {Zener}},\ }\bibfield  {title} {\bibinfo {title} {Non-adiabatic crossing of
  energy levels},\ }\href {https://doi.org/10.1098/rspa.1932.0165} {\bibfield
  {journal} {\bibinfo  {journal} {Proc. R. Soc. Lond. A}\ }\textbf {\bibinfo
  {volume} {137}},\ \bibinfo {pages} {696} (\bibinfo {year}
  {1932})}\BibitemShut {NoStop}%
\bibitem [{\citenamefont {Lloyd}(1996)}]{Ll96}%
  \BibitemOpen
  \bibfield  {author} {\bibinfo {author} {\bibfnamefont {S.}~\bibnamefont
  {Lloyd}},\ }\bibfield  {title} {\bibinfo {title} {{Universal Quantum
  Simulators}},\ }\href {https://doi.org/10.1126/science.273.5278.1073}
  {\bibfield  {journal} {\bibinfo  {journal} {Science}\ }\textbf {\bibinfo
  {volume} {273}},\ \bibinfo {pages} {1073} (\bibinfo {year}
  {1996})}\BibitemShut {NoStop}%
\bibitem [{\citenamefont {Demirplak}\ and\ \citenamefont
  {Rice}(2003)}]{DeRi03}%
  \BibitemOpen
  \bibfield  {author} {\bibinfo {author} {\bibfnamefont {M.}~\bibnamefont
  {Demirplak}}\ and\ \bibinfo {author} {\bibfnamefont {S.~A.}\ \bibnamefont
  {Rice}},\ }\bibfield  {title} {\bibinfo {title} {{Adiabatic Population
  Transfer with Control Fields}},\ }\href {https://doi.org/10.1021/jp030708a}
  {\bibfield  {journal} {\bibinfo  {journal} {J. Phys. Chem. A}\ }\textbf
  {\bibinfo {volume} {107}},\ \bibinfo {pages} {9937} (\bibinfo {year}
  {2003})}\BibitemShut {NoStop}%
\bibitem [{\citenamefont {Demirplak}\ and\ \citenamefont
  {Rice}(2005)}]{DeRi05}%
  \BibitemOpen
  \bibfield  {author} {\bibinfo {author} {\bibfnamefont {M.}~\bibnamefont
  {Demirplak}}\ and\ \bibinfo {author} {\bibfnamefont {S.~A.}\ \bibnamefont
  {Rice}},\ }\bibfield  {title} {\bibinfo {title} {{Assisted Adiabatic Passage
  Revisited}},\ }\href {https://doi.org/10.1021/jp040647w} {\bibfield
  {journal} {\bibinfo  {journal} {J. Phys. Chem. B}\ }\textbf {\bibinfo
  {volume} {109}},\ \bibinfo {pages} {6838} (\bibinfo {year}
  {2005})}\BibitemShut {NoStop}%
\bibitem [{\citenamefont {Berry}(2009)}]{Be09}%
  \BibitemOpen
  \bibfield  {author} {\bibinfo {author} {\bibfnamefont {M.~V.}\ \bibnamefont
  {Berry}},\ }\bibfield  {title} {\bibinfo {title} {Transitionless quantum
  driving},\ }\href {https://doi.org/10.1088/1751-8113/42/36/365303} {\bibfield
   {journal} {\bibinfo  {journal} {J. Phys. A: Math. Theor.}\ }\textbf
  {\bibinfo {volume} {42}},\ \bibinfo {pages} {365303} (\bibinfo {year}
  {2009})}\BibitemShut {NoStop}%
\bibitem [{\citenamefont {Chen}\ \emph {et~al.}(2010)\citenamefont {Chen},
  \citenamefont {Lizuain}, \citenamefont {Ruschhaupt}, \citenamefont
  {Gu\'ery-Odelin},\ and\ \citenamefont {Muga}}]{ChEtAl10}%
  \BibitemOpen
  \bibfield  {author} {\bibinfo {author} {\bibfnamefont {X.}~\bibnamefont
  {Chen}}, \bibinfo {author} {\bibfnamefont {I.}~\bibnamefont {Lizuain}},
  \bibinfo {author} {\bibfnamefont {A.}~\bibnamefont {Ruschhaupt}}, \bibinfo
  {author} {\bibfnamefont {D.}~\bibnamefont {Gu\'ery-Odelin}},\ and\ \bibinfo
  {author} {\bibfnamefont {J.~G.}\ \bibnamefont {Muga}},\ }\bibfield  {title}
  {\bibinfo {title} {{Shortcut to Adiabatic Passage in Two- and Three-Level
  Atoms}},\ }\href {https://doi.org/10.1103/PhysRevLett.105.123003} {\bibfield
  {journal} {\bibinfo  {journal} {Phys. Rev. Lett.}\ }\textbf {\bibinfo
  {volume} {105}},\ \bibinfo {pages} {123003} (\bibinfo {year}
  {2010})}\BibitemShut {NoStop}%
\bibitem [{\citenamefont {del Campo}\ \emph {et~al.}(2012)\citenamefont {del
  Campo}, \citenamefont {Rams},\ and\ \citenamefont {Zurek}}]{DeRaZu12}%
  \BibitemOpen
  \bibfield  {author} {\bibinfo {author} {\bibfnamefont {A.}~\bibnamefont {del
  Campo}}, \bibinfo {author} {\bibfnamefont {M.~M.}\ \bibnamefont {Rams}},\
  and\ \bibinfo {author} {\bibfnamefont {W.~H.}\ \bibnamefont {Zurek}},\
  }\bibfield  {title} {\bibinfo {title} {{Assisted Finite-Rate Adiabatic
  Passage Across a Quantum Critical Point: Exact Solution for the Quantum Ising
  Model}},\ }\href {https://doi.org/10.1103/PhysRevLett.109.115703} {\bibfield
  {journal} {\bibinfo  {journal} {Phys. Rev. Lett.}\ }\textbf {\bibinfo
  {volume} {109}},\ \bibinfo {pages} {115703} (\bibinfo {year}
  {2012})}\BibitemShut {NoStop}%
\bibitem [{\citenamefont {Saberi}\ \emph {et~al.}(2014)\citenamefont {Saberi},
  \citenamefont {Opatrn\'y}, \citenamefont {M\o{}lmer},\ and\ \citenamefont
  {del Campo}}]{SaEtAl14}%
  \BibitemOpen
  \bibfield  {author} {\bibinfo {author} {\bibfnamefont {H.}~\bibnamefont
  {Saberi}}, \bibinfo {author} {\bibfnamefont {T.}~\bibnamefont {Opatrn\'y}},
  \bibinfo {author} {\bibfnamefont {K.}~\bibnamefont {M\o{}lmer}},\ and\
  \bibinfo {author} {\bibfnamefont {A.}~\bibnamefont {del Campo}},\ }\bibfield
  {title} {\bibinfo {title} {Adiabatic tracking of quantum many-body
  dynamics},\ }\href {https://doi.org/10.1103/PhysRevA.90.060301} {\bibfield
  {journal} {\bibinfo  {journal} {Phys. Rev. A}\ }\textbf {\bibinfo {volume}
  {90}},\ \bibinfo {pages} {060301} (\bibinfo {year} {2014})}\BibitemShut
  {NoStop}%
\bibitem [{\citenamefont {Cao}\ \emph {et~al.}(2021)\citenamefont {Cao},
  \citenamefont {Xue}, \citenamefont {Shannon},\ and\ \citenamefont
  {Joynt}}]{ChEtAl21}%
  \BibitemOpen
  \bibfield  {author} {\bibinfo {author} {\bibfnamefont {C.}~\bibnamefont
  {Cao}}, \bibinfo {author} {\bibfnamefont {J.}~\bibnamefont {Xue}}, \bibinfo
  {author} {\bibfnamefont {N.}~\bibnamefont {Shannon}},\ and\ \bibinfo {author}
  {\bibfnamefont {R.}~\bibnamefont {Joynt}},\ }\bibfield  {title} {\bibinfo
  {title} {Speedup of the quantum adiabatic algorithm using delocalization
  catalysis},\ }\href {https://doi.org/10.1103/PhysRevResearch.3.013092}
  {\bibfield  {journal} {\bibinfo  {journal} {Phys. Rev. Res.}\ }\textbf
  {\bibinfo {volume} {3}},\ \bibinfo {pages} {013092} (\bibinfo {year}
  {2021})}\BibitemShut {NoStop}%
\bibitem [{\citenamefont {Sels}\ and\ \citenamefont
  {Polkovnikov}(2017)}]{SePo17}%
  \BibitemOpen
  \bibfield  {author} {\bibinfo {author} {\bibfnamefont {D.}~\bibnamefont
  {Sels}}\ and\ \bibinfo {author} {\bibfnamefont {A.}~\bibnamefont
  {Polkovnikov}},\ }\bibfield  {title} {\bibinfo {title} {Minimizing
  irreversible losses in quantum systems by local counterdiabatic driving},\
  }\href {https://doi.org/10.1073/pnas.1619826114} {\bibfield  {journal}
  {\bibinfo  {journal} {Proc. Natl. Acad. Sci. U.S.A.}\ }\textbf {\bibinfo
  {volume} {114}},\ \bibinfo {pages} {E3909} (\bibinfo {year}
  {2017})}\BibitemShut {NoStop}%
\bibitem [{\citenamefont {Kolodrubetz}\ \emph {et~al.}(2017)\citenamefont
  {Kolodrubetz}, \citenamefont {Sels}, \citenamefont {Mehta},\ and\
  \citenamefont {Polkovnikov}}]{KoEtAl17}%
  \BibitemOpen
  \bibfield  {author} {\bibinfo {author} {\bibfnamefont {M.}~\bibnamefont
  {Kolodrubetz}}, \bibinfo {author} {\bibfnamefont {D.}~\bibnamefont {Sels}},
  \bibinfo {author} {\bibfnamefont {P.}~\bibnamefont {Mehta}},\ and\ \bibinfo
  {author} {\bibfnamefont {A.}~\bibnamefont {Polkovnikov}},\ }\bibfield
  {title} {\bibinfo {title} {Geometry and non-adiabatic response in quantum and
  classical systems},\ }\href
  {https://doi.org/https://doi.org/10.1016/j.physrep.2017.07.001} {\bibfield
  {journal} {\bibinfo  {journal} {Phys. Rep.}\ }\textbf {\bibinfo {volume}
  {697}},\ \bibinfo {pages} {1} (\bibinfo {year} {2017})}\BibitemShut {NoStop}%
\bibitem [{\citenamefont {Hegade}\ \emph
  {et~al.}(2021{\natexlab{a}})\citenamefont {Hegade}, \citenamefont {Paul},
  \citenamefont {Ding}, \citenamefont {Sanz}, \citenamefont
  {Albarr\'an-Arriagada}, \citenamefont {Solano},\ and\ \citenamefont
  {Chen}}]{HeEtAl21a}%
  \BibitemOpen
  \bibfield  {author} {\bibinfo {author} {\bibfnamefont {N.~N.}\ \bibnamefont
  {Hegade}}, \bibinfo {author} {\bibfnamefont {K.}~\bibnamefont {Paul}},
  \bibinfo {author} {\bibfnamefont {Y.}~\bibnamefont {Ding}}, \bibinfo {author}
  {\bibfnamefont {M.}~\bibnamefont {Sanz}}, \bibinfo {author} {\bibfnamefont
  {F.}~\bibnamefont {Albarr\'an-Arriagada}}, \bibinfo {author} {\bibfnamefont
  {E.}~\bibnamefont {Solano}},\ and\ \bibinfo {author} {\bibfnamefont
  {X.}~\bibnamefont {Chen}},\ }\bibfield  {title} {\bibinfo {title} {{Shortcuts
  to Adiabaticity in Digitized Adiabatic Quantum Computing}},\ }\href
  {https://doi.org/10.1103/PhysRevApplied.15.024038} {\bibfield  {journal}
  {\bibinfo  {journal} {Phys. Rev. Appl.}\ }\textbf {\bibinfo {volume} {15}},\
  \bibinfo {pages} {024038} (\bibinfo {year} {2021}{\natexlab{a}})}\BibitemShut
  {NoStop}%
\bibitem [{\citenamefont {Hegade}\ \emph
  {et~al.}(2021{\natexlab{b}})\citenamefont {Hegade}, \citenamefont {Paul},
  \citenamefont {Albarr\'an-Arriagada}, \citenamefont {Chen},\ and\
  \citenamefont {Solano}}]{HeEtAl21b}%
  \BibitemOpen
  \bibfield  {author} {\bibinfo {author} {\bibfnamefont {N.~N.}\ \bibnamefont
  {Hegade}}, \bibinfo {author} {\bibfnamefont {K.}~\bibnamefont {Paul}},
  \bibinfo {author} {\bibfnamefont {F.}~\bibnamefont {Albarr\'an-Arriagada}},
  \bibinfo {author} {\bibfnamefont {X.}~\bibnamefont {Chen}},\ and\ \bibinfo
  {author} {\bibfnamefont {E.}~\bibnamefont {Solano}},\ }\bibfield  {title}
  {\bibinfo {title} {Digitized adiabatic quantum factorization},\ }\href
  {https://doi.org/10.1103/PhysRevA.104.L050403} {\bibfield  {journal}
  {\bibinfo  {journal} {Phys. Rev. A}\ }\textbf {\bibinfo {volume} {104}},\
  \bibinfo {pages} {L050403} (\bibinfo {year}
  {2021}{\natexlab{b}})}\BibitemShut {NoStop}%
\bibitem [{\citenamefont {Hegade}\ \emph
  {et~al.}(2022{\natexlab{a}})\citenamefont {Hegade}, \citenamefont {Chen},\
  and\ \citenamefont {Solano}}]{HeChSo22}%
  \BibitemOpen
  \bibfield  {author} {\bibinfo {author} {\bibfnamefont {N.~N.}\ \bibnamefont
  {Hegade}}, \bibinfo {author} {\bibfnamefont {X.}~\bibnamefont {Chen}},\ and\
  \bibinfo {author} {\bibfnamefont {E.}~\bibnamefont {Solano}},\ }\bibfield
  {title} {\bibinfo {title} {Digitized counterdiabatic quantum optimization},\
  }\href {https://doi.org/10.1103/PhysRevResearch.4.L042030} {\bibfield
  {journal} {\bibinfo  {journal} {Phys. Rev. Res.}\ }\textbf {\bibinfo {volume}
  {4}},\ \bibinfo {pages} {L042030} (\bibinfo {year}
  {2022}{\natexlab{a}})}\BibitemShut {NoStop}%
\bibitem [{\citenamefont {Hegade}\ \emph
  {et~al.}(2022{\natexlab{b}})\citenamefont {Hegade}, \citenamefont
  {Chandarana}, \citenamefont {Paul}, \citenamefont {Chen}, \citenamefont
  {Albarr\'an-Arriagada},\ and\ \citenamefont {Solano}}]{HeEtAl22A}%
  \BibitemOpen
  \bibfield  {author} {\bibinfo {author} {\bibfnamefont {N.~N.}\ \bibnamefont
  {Hegade}}, \bibinfo {author} {\bibfnamefont {P.}~\bibnamefont {Chandarana}},
  \bibinfo {author} {\bibfnamefont {K.}~\bibnamefont {Paul}}, \bibinfo {author}
  {\bibfnamefont {X.}~\bibnamefont {Chen}}, \bibinfo {author} {\bibfnamefont
  {F.}~\bibnamefont {Albarr\'an-Arriagada}},\ and\ \bibinfo {author}
  {\bibfnamefont {E.}~\bibnamefont {Solano}},\ }\bibfield  {title} {\bibinfo
  {title} {Portfolio optimization with digitized counterdiabatic quantum
  algorithms},\ }\href {https://doi.org/10.1103/PhysRevResearch.4.043204}
  {\bibfield  {journal} {\bibinfo  {journal} {Phys. Rev. Res.}\ }\textbf
  {\bibinfo {volume} {4}},\ \bibinfo {pages} {043204} (\bibinfo {year}
  {2022}{\natexlab{b}})}\BibitemShut {NoStop}%
\bibitem [{\citenamefont {Verstraete}\ \emph {et~al.}(2008)\citenamefont
  {Verstraete}, \citenamefont {Murg},\ and\ \citenamefont {Cirac}}]{VeMuCi08}%
  \BibitemOpen
  \bibfield  {author} {\bibinfo {author} {\bibfnamefont {F.}~\bibnamefont
  {Verstraete}}, \bibinfo {author} {\bibfnamefont {V.}~\bibnamefont {Murg}},\
  and\ \bibinfo {author} {\bibfnamefont {J.~I.}\ \bibnamefont {Cirac}},\
  }\bibfield  {title} {\bibinfo {title} {Matrix product states, projected
  entangled pair states, and variational renormalization group methods for
  quantum spin systems},\ }\href {https://doi.org/10.1080/14789940801912366}
  {\bibfield  {journal} {\bibinfo  {journal} {Adv. Phys.}\ }\textbf {\bibinfo
  {volume} {57}},\ \bibinfo {pages} {143} (\bibinfo {year} {2008})}\BibitemShut
  {NoStop}%
\bibitem [{\citenamefont {Orús}(2014)}]{Or14}%
  \BibitemOpen
  \bibfield  {author} {\bibinfo {author} {\bibfnamefont {R.}~\bibnamefont
  {Orús}},\ }\bibfield  {title} {\bibinfo {title} {A practical introduction to
  tensor networks: {Matrix} product states and projected entangled pair
  states},\ }\href {https://doi.org/https://doi.org/10.1016/j.aop.2014.06.013}
  {\bibfield  {journal} {\bibinfo  {journal} {Ann. Phys.}\ }\textbf {\bibinfo
  {volume} {349}},\ \bibinfo {pages} {117} (\bibinfo {year}
  {2014})}\BibitemShut {NoStop}%
\bibitem [{\citenamefont {Ba\~{n}uls}(2023)}]{Ba23}%
  \BibitemOpen
  \bibfield  {author} {\bibinfo {author} {\bibfnamefont {M.~C.}\ \bibnamefont
  {Ba\~{n}uls}},\ }\bibfield  {title} {\bibinfo {title} {{Tensor Network
  Algorithms: A Route Map}},\ }\href
  {https://doi.org/10.1146/annurev-conmatphys-040721-022705} {\bibfield
  {journal} {\bibinfo  {journal} {Annu. Rev. Condens. Matter Phys.}\ }\textbf
  {\bibinfo {volume} {14}},\ \bibinfo {pages} {173} (\bibinfo {year}
  {2023})}\BibitemShut {NoStop}%
\bibitem [{\citenamefont {Benedetti}\ \emph {et~al.}(2019)\citenamefont
  {Benedetti}, \citenamefont {Lloyd}, \citenamefont {Sack},\ and\ \citenamefont
  {Fiorentini}}]{BeEtAl19}%
  \BibitemOpen
  \bibfield  {author} {\bibinfo {author} {\bibfnamefont {M.}~\bibnamefont
  {Benedetti}}, \bibinfo {author} {\bibfnamefont {E.}~\bibnamefont {Lloyd}},
  \bibinfo {author} {\bibfnamefont {S.}~\bibnamefont {Sack}},\ and\ \bibinfo
  {author} {\bibfnamefont {M.}~\bibnamefont {Fiorentini}},\ }\bibfield  {title}
  {\bibinfo {title} {Parameterized quantum circuits as machine learning
  models},\ }\href {https://doi.org/10.1088/2058-9565/ab4eb5} {\bibfield
  {journal} {\bibinfo  {journal} {Quantum Sci. Technol.}\ }\textbf {\bibinfo
  {volume} {4}},\ \bibinfo {pages} {043001} (\bibinfo {year}
  {2019})}\BibitemShut {NoStop}%
\bibitem [{\citenamefont {Cerezo}\ \emph
  {et~al.}(2021{\natexlab{a}})\citenamefont {Cerezo}, \citenamefont
  {Arrasmith}, \citenamefont {Babbush}, \citenamefont {Benjamin}, \citenamefont
  {Endo}, \citenamefont {Fujii}, \citenamefont {McClean}, \citenamefont
  {Mitarai}, \citenamefont {Yuan}, \citenamefont {Cincio},\ and\ \citenamefont
  {Coles}}]{CeEtAl21}%
  \BibitemOpen
  \bibfield  {author} {\bibinfo {author} {\bibfnamefont {M.}~\bibnamefont
  {Cerezo}}, \bibinfo {author} {\bibfnamefont {A.}~\bibnamefont {Arrasmith}},
  \bibinfo {author} {\bibfnamefont {R.}~\bibnamefont {Babbush}}, \bibinfo
  {author} {\bibfnamefont {S.~C.}\ \bibnamefont {Benjamin}}, \bibinfo {author}
  {\bibfnamefont {S.}~\bibnamefont {Endo}}, \bibinfo {author} {\bibfnamefont
  {K.}~\bibnamefont {Fujii}}, \bibinfo {author} {\bibfnamefont {J.~R.}\
  \bibnamefont {McClean}}, \bibinfo {author} {\bibfnamefont {K.}~\bibnamefont
  {Mitarai}}, \bibinfo {author} {\bibfnamefont {X.}~\bibnamefont {Yuan}},
  \bibinfo {author} {\bibfnamefont {L.}~\bibnamefont {Cincio}},\ and\ \bibinfo
  {author} {\bibfnamefont {P.~J.}\ \bibnamefont {Coles}},\ }\bibfield  {title}
  {\bibinfo {title} {Variational quantum algorithms},\ }\href
  {https://doi.org/10.1038/s42254-021-00348-9} {\bibfield  {journal} {\bibinfo
  {journal} {Nat. Rev. Phys.}\ }\textbf {\bibinfo {volume} {3}},\ \bibinfo
  {pages} {625} (\bibinfo {year} {2021}{\natexlab{a}})}\BibitemShut {NoStop}%
\bibitem [{\citenamefont {Bharti}\ \emph {et~al.}(2022)\citenamefont {Bharti},
  \citenamefont {Cervera-Lierta}, \citenamefont {Kyaw}, \citenamefont {Haug},
  \citenamefont {Alperin-Lea}, \citenamefont {Anand}, \citenamefont {Degroote},
  \citenamefont {Heimonen}, \citenamefont {Kottmann}, \citenamefont {Menke},
  \citenamefont {Mok}, \citenamefont {Sim}, \citenamefont {Kwek},\ and\
  \citenamefont {Aspuru-Guzik}}]{BhEtAl22}%
  \BibitemOpen
  \bibfield  {author} {\bibinfo {author} {\bibfnamefont {K.}~\bibnamefont
  {Bharti}}, \bibinfo {author} {\bibfnamefont {A.}~\bibnamefont
  {Cervera-Lierta}}, \bibinfo {author} {\bibfnamefont {T.~H.}\ \bibnamefont
  {Kyaw}}, \bibinfo {author} {\bibfnamefont {T.}~\bibnamefont {Haug}}, \bibinfo
  {author} {\bibfnamefont {S.}~\bibnamefont {Alperin-Lea}}, \bibinfo {author}
  {\bibfnamefont {A.}~\bibnamefont {Anand}}, \bibinfo {author} {\bibfnamefont
  {M.}~\bibnamefont {Degroote}}, \bibinfo {author} {\bibfnamefont
  {H.}~\bibnamefont {Heimonen}}, \bibinfo {author} {\bibfnamefont {J.~S.}\
  \bibnamefont {Kottmann}}, \bibinfo {author} {\bibfnamefont {T.}~\bibnamefont
  {Menke}}, \bibinfo {author} {\bibfnamefont {W.-K.}\ \bibnamefont {Mok}},
  \bibinfo {author} {\bibfnamefont {S.}~\bibnamefont {Sim}}, \bibinfo {author}
  {\bibfnamefont {L.-C.}\ \bibnamefont {Kwek}},\ and\ \bibinfo {author}
  {\bibfnamefont {A.}~\bibnamefont {Aspuru-Guzik}},\ }\bibfield  {title}
  {\bibinfo {title} {Noisy intermediate-scale quantum algorithms},\ }\href
  {https://doi.org/10.1103/RevModPhys.94.015004} {\bibfield  {journal}
  {\bibinfo  {journal} {Rev. Mod. Phys.}\ }\textbf {\bibinfo {volume} {94}},\
  \bibinfo {pages} {015004} (\bibinfo {year} {2022})}\BibitemShut {NoStop}%
\bibitem [{\citenamefont {Mansuroglu}\ \emph {et~al.}(2023)\citenamefont
  {Mansuroglu}, \citenamefont {Eckstein}, \citenamefont {Nützel},
  \citenamefont {Wilkinson},\ and\ \citenamefont {Hartmann}}]{MaEtAl23}%
  \BibitemOpen
  \bibfield  {author} {\bibinfo {author} {\bibfnamefont {R.}~\bibnamefont
  {Mansuroglu}}, \bibinfo {author} {\bibfnamefont {T.}~\bibnamefont
  {Eckstein}}, \bibinfo {author} {\bibfnamefont {L.}~\bibnamefont {Nützel}},
  \bibinfo {author} {\bibfnamefont {S.~A.}\ \bibnamefont {Wilkinson}},\ and\
  \bibinfo {author} {\bibfnamefont {M.~J.}\ \bibnamefont {Hartmann}},\
  }\bibfield  {title} {\bibinfo {title} {{Variational Hamiltonian} simulation
  for translational invariant systems via classical pre-processing},\ }\href
  {https://doi.org/10.1088/2058-9565/acb1d0} {\bibfield  {journal} {\bibinfo
  {journal} {Quantum Sci. Technol.}\ }\textbf {\bibinfo {volume} {8}},\
  \bibinfo {pages} {025006} (\bibinfo {year} {2023})}\BibitemShut {NoStop}%
\bibitem [{\citenamefont {Tepaske}\ \emph {et~al.}(2023)\citenamefont
  {Tepaske}, \citenamefont {Hahn},\ and\ \citenamefont {Luitz}}]{TeHaLu23}%
  \BibitemOpen
  \bibfield  {author} {\bibinfo {author} {\bibfnamefont {M.~S.~J.}\
  \bibnamefont {Tepaske}}, \bibinfo {author} {\bibfnamefont {D.}~\bibnamefont
  {Hahn}},\ and\ \bibinfo {author} {\bibfnamefont {D.~J.}\ \bibnamefont
  {Luitz}},\ }\bibfield  {title} {\bibinfo {title} {Optimal compression of
  quantum many-body time evolution operators into brickwall circuits},\ }\href
  {https://doi.org/10.21468/SciPostPhys.14.4.073} {\bibfield  {journal}
  {\bibinfo  {journal} {SciPost Phys.}\ }\textbf {\bibinfo {volume} {14}},\
  \bibinfo {pages} {073} (\bibinfo {year} {2023})}\BibitemShut {NoStop}%
\bibitem [{\citenamefont {Mc~Keever}\ and\ \citenamefont
  {Lubasch}(2023)}]{McLu23}%
  \BibitemOpen
  \bibfield  {author} {\bibinfo {author} {\bibfnamefont {C.}~\bibnamefont
  {Mc~Keever}}\ and\ \bibinfo {author} {\bibfnamefont {M.}~\bibnamefont
  {Lubasch}},\ }\bibfield  {title} {\bibinfo {title} {{Classically optimized
  Hamiltonian simulation}},\ }\href
  {https://doi.org/10.1103/PhysRevResearch.5.023146} {\bibfield  {journal}
  {\bibinfo  {journal} {Phys. Rev. Res.}\ }\textbf {\bibinfo {volume} {5}},\
  \bibinfo {pages} {023146} (\bibinfo {year} {2023})}\BibitemShut {NoStop}%
\bibitem [{\citenamefont {Pandey}\ \emph {et~al.}(2020)\citenamefont {Pandey},
  \citenamefont {Claeys}, \citenamefont {Campbell}, \citenamefont
  {Polkovnikov},\ and\ \citenamefont {Sels}}]{Pandey2020}%
  \BibitemOpen
  \bibfield  {author} {\bibinfo {author} {\bibfnamefont {M.}~\bibnamefont
  {Pandey}}, \bibinfo {author} {\bibfnamefont {P.~W.}\ \bibnamefont {Claeys}},
  \bibinfo {author} {\bibfnamefont {D.~K.}\ \bibnamefont {Campbell}}, \bibinfo
  {author} {\bibfnamefont {A.}~\bibnamefont {Polkovnikov}},\ and\ \bibinfo
  {author} {\bibfnamefont {D.}~\bibnamefont {Sels}},\ }\bibfield  {title}
  {\bibinfo {title} {{Adiabatic Eigenstate Deformations as a Sensitive Probe
  for Quantum Chaos}},\ }\href {https://doi.org/10.1103/PhysRevX.10.041017}
  {\bibfield  {journal} {\bibinfo  {journal} {Phys. Rev. X}\ }\textbf {\bibinfo
  {volume} {10}},\ \bibinfo {pages} {041017} (\bibinfo {year}
  {2020})}\BibitemShut {NoStop}%
\bibitem [{\citenamefont {Verstraete}\ \emph {et~al.}(2004)\citenamefont
  {Verstraete}, \citenamefont {Garc\'{\i}a-Ripoll},\ and\ \citenamefont
  {Cirac}}]{VeGaCi04}%
  \BibitemOpen
  \bibfield  {author} {\bibinfo {author} {\bibfnamefont {F.}~\bibnamefont
  {Verstraete}}, \bibinfo {author} {\bibfnamefont {J.~J.}\ \bibnamefont
  {Garc\'{\i}a-Ripoll}},\ and\ \bibinfo {author} {\bibfnamefont {J.~I.}\
  \bibnamefont {Cirac}},\ }\bibfield  {title} {\bibinfo {title} {{Matrix
  Product Density Operators: Simulation of Finite-Temperature and Dissipative
  Systems}},\ }\href {https://doi.org/10.1103/PhysRevLett.93.207204} {\bibfield
   {journal} {\bibinfo  {journal} {Phys. Rev. Lett.}\ }\textbf {\bibinfo
  {volume} {93}},\ \bibinfo {pages} {207204} (\bibinfo {year}
  {2004})}\BibitemShut {NoStop}%
\bibitem [{\citenamefont {Zwolak}\ and\ \citenamefont {Vidal}(2004)}]{ZwVi04}%
  \BibitemOpen
  \bibfield  {author} {\bibinfo {author} {\bibfnamefont {M.}~\bibnamefont
  {Zwolak}}\ and\ \bibinfo {author} {\bibfnamefont {G.}~\bibnamefont {Vidal}},\
  }\bibfield  {title} {\bibinfo {title} {{Mixed-State Dynamics in
  One-Dimensional Quantum Lattice Systems: A Time-Dependent Superoperator
  Renormalization Algorithm}},\ }\href
  {https://doi.org/10.1103/PhysRevLett.93.207205} {\bibfield  {journal}
  {\bibinfo  {journal} {Phys. Rev. Lett.}\ }\textbf {\bibinfo {volume} {93}},\
  \bibinfo {pages} {207205} (\bibinfo {year} {2004})}\BibitemShut {NoStop}%
\bibitem [{\citenamefont {Claeys}\ \emph {et~al.}(2019)\citenamefont {Claeys},
  \citenamefont {Pandey}, \citenamefont {Sels},\ and\ \citenamefont
  {Polkovnikov}}]{ClEtAl19}%
  \BibitemOpen
  \bibfield  {author} {\bibinfo {author} {\bibfnamefont {P.~W.}\ \bibnamefont
  {Claeys}}, \bibinfo {author} {\bibfnamefont {M.}~\bibnamefont {Pandey}},
  \bibinfo {author} {\bibfnamefont {D.}~\bibnamefont {Sels}},\ and\ \bibinfo
  {author} {\bibfnamefont {A.}~\bibnamefont {Polkovnikov}},\ }\bibfield
  {title} {\bibinfo {title} {{Floquet-Engineering Counterdiabatic Protocols in
  Quantum Many-Body Systems}},\ }\href
  {https://doi.org/10.1103/PhysRevLett.123.090602} {\bibfield  {journal}
  {\bibinfo  {journal} {Phys. Rev. Lett.}\ }\textbf {\bibinfo {volume} {123}},\
  \bibinfo {pages} {090602} (\bibinfo {year} {2019})}\BibitemShut {NoStop}%
\bibitem [{\citenamefont {Mazzola}(2024)}]{Ma23}%
  \BibitemOpen
  \bibfield  {author} {\bibinfo {author} {\bibfnamefont {G.}~\bibnamefont
  {Mazzola}},\ }\bibfield  {title} {\bibinfo {title} {Quantum computing for
  chemistry and physics applications from a {Monte Carlo} perspective},\ }\href
  {https://doi.org/10.1063/5.0173591} {\bibfield  {journal} {\bibinfo
  {journal} {J. Chem. Phys.}\ }\textbf {\bibinfo {volume} {160}},\ \bibinfo
  {pages} {010901} (\bibinfo {year} {2024})}\BibitemShut {NoStop}%
\bibitem [{\citenamefont {Li}\ and\ \citenamefont {Benjamin}(2017)}]{LiBe17}%
  \BibitemOpen
  \bibfield  {author} {\bibinfo {author} {\bibfnamefont {Y.}~\bibnamefont
  {Li}}\ and\ \bibinfo {author} {\bibfnamefont {S.~C.}\ \bibnamefont
  {Benjamin}},\ }\bibfield  {title} {\bibinfo {title} {{Efficient Variational
  Quantum Simulator Incorporating Active Error Minimization}},\ }\href
  {https://doi.org/10.1103/PhysRevX.7.021050} {\bibfield  {journal} {\bibinfo
  {journal} {Phys. Rev. X}\ }\textbf {\bibinfo {volume} {7}},\ \bibinfo {pages}
  {021050} (\bibinfo {year} {2017})}\BibitemShut {NoStop}%
\bibitem [{\citenamefont {Yuan}\ \emph {et~al.}(2019)\citenamefont {Yuan},
  \citenamefont {Endo}, \citenamefont {Zhao}, \citenamefont {Li},\ and\
  \citenamefont {Benjamin}}]{YuEtAl19}%
  \BibitemOpen
  \bibfield  {author} {\bibinfo {author} {\bibfnamefont {X.}~\bibnamefont
  {Yuan}}, \bibinfo {author} {\bibfnamefont {S.}~\bibnamefont {Endo}}, \bibinfo
  {author} {\bibfnamefont {Q.}~\bibnamefont {Zhao}}, \bibinfo {author}
  {\bibfnamefont {Y.}~\bibnamefont {Li}},\ and\ \bibinfo {author}
  {\bibfnamefont {S.~C.}\ \bibnamefont {Benjamin}},\ }\bibfield  {title}
  {\bibinfo {title} {Theory of variational quantum simulation},\ }\href
  {https://doi.org/10.22331/q-2019-10-07-191} {\bibfield  {journal} {\bibinfo
  {journal} {{Quantum}}\ }\textbf {\bibinfo {volume} {3}},\ \bibinfo {pages}
  {191} (\bibinfo {year} {2019})}\BibitemShut {NoStop}%
\bibitem [{\citenamefont {Benedetti}\ \emph {et~al.}(2021)\citenamefont
  {Benedetti}, \citenamefont {Fiorentini},\ and\ \citenamefont
  {Lubasch}}]{BeFiLu21}%
  \BibitemOpen
  \bibfield  {author} {\bibinfo {author} {\bibfnamefont {M.}~\bibnamefont
  {Benedetti}}, \bibinfo {author} {\bibfnamefont {M.}~\bibnamefont
  {Fiorentini}},\ and\ \bibinfo {author} {\bibfnamefont {M.}~\bibnamefont
  {Lubasch}},\ }\bibfield  {title} {\bibinfo {title} {Hardware-efficient
  variational quantum algorithms for time evolution},\ }\href
  {https://doi.org/10.1103/PhysRevResearch.3.033083} {\bibfield  {journal}
  {\bibinfo  {journal} {Phys. Rev. Res.}\ }\textbf {\bibinfo {volume} {3}},\
  \bibinfo {pages} {033083} (\bibinfo {year} {2021})}\BibitemShut {NoStop}%
\bibitem [{\citenamefont {Barison}\ \emph {et~al.}(2021)\citenamefont
  {Barison}, \citenamefont {Vicentini},\ and\ \citenamefont
  {Carleo}}]{BaViCa21}%
  \BibitemOpen
  \bibfield  {author} {\bibinfo {author} {\bibfnamefont {S.}~\bibnamefont
  {Barison}}, \bibinfo {author} {\bibfnamefont {F.}~\bibnamefont {Vicentini}},\
  and\ \bibinfo {author} {\bibfnamefont {G.}~\bibnamefont {Carleo}},\
  }\bibfield  {title} {\bibinfo {title} {An efficient quantum algorithm for the
  time evolution of parameterized circuits},\ }\href
  {https://doi.org/10.22331/q-2021-07-28-512} {\bibfield  {journal} {\bibinfo
  {journal} {{Quantum}}\ }\textbf {\bibinfo {volume} {5}},\ \bibinfo {pages}
  {512} (\bibinfo {year} {2021})}\BibitemShut {NoStop}%
\bibitem [{\citenamefont {Chandarana}\ \emph {et~al.}(2022)\citenamefont
  {Chandarana}, \citenamefont {Hegade}, \citenamefont {Paul}, \citenamefont
  {Albarr\'an-Arriagada}, \citenamefont {Solano}, \citenamefont {del Campo},\
  and\ \citenamefont {Chen}}]{ChEtAl22}%
  \BibitemOpen
  \bibfield  {author} {\bibinfo {author} {\bibfnamefont {P.}~\bibnamefont
  {Chandarana}}, \bibinfo {author} {\bibfnamefont {N.~N.}\ \bibnamefont
  {Hegade}}, \bibinfo {author} {\bibfnamefont {K.}~\bibnamefont {Paul}},
  \bibinfo {author} {\bibfnamefont {F.}~\bibnamefont {Albarr\'an-Arriagada}},
  \bibinfo {author} {\bibfnamefont {E.}~\bibnamefont {Solano}}, \bibinfo
  {author} {\bibfnamefont {A.}~\bibnamefont {del Campo}},\ and\ \bibinfo
  {author} {\bibfnamefont {X.}~\bibnamefont {Chen}},\ }\bibfield  {title}
  {\bibinfo {title} {Digitized-counterdiabatic quantum approximate optimization
  algorithm},\ }\href {https://doi.org/10.1103/PhysRevResearch.4.013141}
  {\bibfield  {journal} {\bibinfo  {journal} {Phys. Rev. Res.}\ }\textbf
  {\bibinfo {volume} {4}},\ \bibinfo {pages} {013141} (\bibinfo {year}
  {2022})}\BibitemShut {NoStop}%
\bibitem [{\citenamefont {Chandarana}\ \emph
  {et~al.}(2023{\natexlab{a}})\citenamefont {Chandarana}, \citenamefont
  {Hegade}, \citenamefont {Montalban}, \citenamefont {Solano},\ and\
  \citenamefont {Chen}}]{ChEtAl23a}%
  \BibitemOpen
  \bibfield  {author} {\bibinfo {author} {\bibfnamefont {P.}~\bibnamefont
  {Chandarana}}, \bibinfo {author} {\bibfnamefont {N.~N.}\ \bibnamefont
  {Hegade}}, \bibinfo {author} {\bibfnamefont {I.}~\bibnamefont {Montalban}},
  \bibinfo {author} {\bibfnamefont {E.}~\bibnamefont {Solano}},\ and\ \bibinfo
  {author} {\bibfnamefont {X.}~\bibnamefont {Chen}},\ }\bibfield  {title}
  {\bibinfo {title} {{Digitized Counterdiabatic Quantum Algorithm for Protein
  Folding}},\ }\href {https://doi.org/10.1103/PhysRevApplied.20.014024}
  {\bibfield  {journal} {\bibinfo  {journal} {Phys. Rev. Appl.}\ }\textbf
  {\bibinfo {volume} {20}},\ \bibinfo {pages} {014024} (\bibinfo {year}
  {2023}{\natexlab{a}})}\BibitemShut {NoStop}%
\bibitem [{\citenamefont {Chandarana}\ \emph
  {et~al.}(2023{\natexlab{b}})\citenamefont {Chandarana}, \citenamefont
  {Vieites}, \citenamefont {Hegade}, \citenamefont {Solano}, \citenamefont
  {Ban},\ and\ \citenamefont {Chen}}]{ChEtAl23b}%
  \BibitemOpen
  \bibfield  {author} {\bibinfo {author} {\bibfnamefont {P.}~\bibnamefont
  {Chandarana}}, \bibinfo {author} {\bibfnamefont {P.~S.}\ \bibnamefont
  {Vieites}}, \bibinfo {author} {\bibfnamefont {N.~N.}\ \bibnamefont {Hegade}},
  \bibinfo {author} {\bibfnamefont {E.}~\bibnamefont {Solano}}, \bibinfo
  {author} {\bibfnamefont {Y.}~\bibnamefont {Ban}},\ and\ \bibinfo {author}
  {\bibfnamefont {X.}~\bibnamefont {Chen}},\ }\bibfield  {title} {\bibinfo
  {title} {Meta-learning digitized-counterdiabatic quantum optimization},\
  }\href {https://doi.org/10.1088/2058-9565/ace54a} {\bibfield  {journal}
  {\bibinfo  {journal} {Quantum Sci. Technol.}\ }\textbf {\bibinfo {volume}
  {8}},\ \bibinfo {pages} {045007} (\bibinfo {year}
  {2023}{\natexlab{b}})}\BibitemShut {NoStop}%
\bibitem [{\citenamefont {Malla}\ \emph {et~al.}(2024)\citenamefont {Malla},
  \citenamefont {Sukeno}, \citenamefont {Yu}, \citenamefont {Wei},
  \citenamefont {Weichselbaum},\ and\ \citenamefont {Konik}}]{MaEtAl24}%
  \BibitemOpen
  \bibfield  {author} {\bibinfo {author} {\bibfnamefont {R.~K.}\ \bibnamefont
  {Malla}}, \bibinfo {author} {\bibfnamefont {H.}~\bibnamefont {Sukeno}},
  \bibinfo {author} {\bibfnamefont {H.}~\bibnamefont {Yu}}, \bibinfo {author}
  {\bibfnamefont {T.-C.}\ \bibnamefont {Wei}}, \bibinfo {author} {\bibfnamefont
  {A.}~\bibnamefont {Weichselbaum}},\ and\ \bibinfo {author} {\bibfnamefont
  {R.~M.}\ \bibnamefont {Konik}},\ }\href@noop {} {\bibinfo {title}
  {{Feedback-based Quantum Algorithm Inspired by Counterdiabatic Driving}}}
  (\bibinfo {year} {2024}),\ \Eprint {https://arxiv.org/abs/2401.15303}
  {arXiv:2401.15303 [quant-ph]} \BibitemShut {NoStop}%
\bibitem [{\citenamefont {McClean}\ \emph {et~al.}(2018)\citenamefont
  {McClean}, \citenamefont {Boixo}, \citenamefont {Smelyanskiy}, \citenamefont
  {Babbush},\ and\ \citenamefont {Neven}}]{McEtAl18}%
  \BibitemOpen
  \bibfield  {author} {\bibinfo {author} {\bibfnamefont {J.~R.}\ \bibnamefont
  {McClean}}, \bibinfo {author} {\bibfnamefont {S.}~\bibnamefont {Boixo}},
  \bibinfo {author} {\bibfnamefont {V.~N.}\ \bibnamefont {Smelyanskiy}},
  \bibinfo {author} {\bibfnamefont {R.}~\bibnamefont {Babbush}},\ and\ \bibinfo
  {author} {\bibfnamefont {H.}~\bibnamefont {Neven}},\ }\bibfield  {title}
  {\bibinfo {title} {Barren plateaus in quantum neural network training
  landscapes},\ }\href {https://doi.org/10.1038/s41467-018-07090-4} {\bibfield
  {journal} {\bibinfo  {journal} {Nat. Commun.}\ }\textbf {\bibinfo {volume}
  {9}},\ \bibinfo {pages} {4812} (\bibinfo {year} {2018})}\BibitemShut
  {NoStop}%
\bibitem [{\citenamefont {Cerezo}\ \emph
  {et~al.}(2021{\natexlab{b}})\citenamefont {Cerezo}, \citenamefont {Sone},
  \citenamefont {Volkoff}, \citenamefont {Cincio},\ and\ \citenamefont
  {Coles}}]{CeEtAl21B}%
  \BibitemOpen
  \bibfield  {author} {\bibinfo {author} {\bibfnamefont {M.}~\bibnamefont
  {Cerezo}}, \bibinfo {author} {\bibfnamefont {A.}~\bibnamefont {Sone}},
  \bibinfo {author} {\bibfnamefont {T.}~\bibnamefont {Volkoff}}, \bibinfo
  {author} {\bibfnamefont {L.}~\bibnamefont {Cincio}},\ and\ \bibinfo {author}
  {\bibfnamefont {P.~J.}\ \bibnamefont {Coles}},\ }\bibfield  {title} {\bibinfo
  {title} {Cost function dependent barren plateaus in shallow parametrized
  quantum circuits},\ }\href {https://doi.org/10.1038/s41467-021-21728-w}
  {\bibfield  {journal} {\bibinfo  {journal} {Nat. Commun.}\ }\textbf {\bibinfo
  {volume} {12}},\ \bibinfo {pages} {1791} (\bibinfo {year}
  {2021}{\natexlab{b}})}\BibitemShut {NoStop}%
\bibitem [{\citenamefont {Holmes}\ \emph {et~al.}(2021)\citenamefont {Holmes},
  \citenamefont {Arrasmith}, \citenamefont {Yan}, \citenamefont {Coles},
  \citenamefont {Albrecht},\ and\ \citenamefont {Sornborger}}]{Holmes2021}%
  \BibitemOpen
  \bibfield  {author} {\bibinfo {author} {\bibfnamefont {Z.}~\bibnamefont
  {Holmes}}, \bibinfo {author} {\bibfnamefont {A.}~\bibnamefont {Arrasmith}},
  \bibinfo {author} {\bibfnamefont {B.}~\bibnamefont {Yan}}, \bibinfo {author}
  {\bibfnamefont {P.~J.}\ \bibnamefont {Coles}}, \bibinfo {author}
  {\bibfnamefont {A.}~\bibnamefont {Albrecht}},\ and\ \bibinfo {author}
  {\bibfnamefont {A.~T.}\ \bibnamefont {Sornborger}},\ }\bibfield  {title}
  {\bibinfo {title} {{Barren Plateaus Preclude Learning Scramblers}},\ }\href
  {https://doi.org/10.1103/PhysRevLett.126.190501} {\bibfield  {journal}
  {\bibinfo  {journal} {Phys. Rev. Lett.}\ }\textbf {\bibinfo {volume} {126}},\
  \bibinfo {pages} {190501} (\bibinfo {year} {2021})}\BibitemShut {NoStop}%
\bibitem [{\citenamefont {Holmes}\ \emph {et~al.}(2022)\citenamefont {Holmes},
  \citenamefont {Sharma}, \citenamefont {Cerezo},\ and\ \citenamefont
  {Coles}}]{Holmes2022}%
  \BibitemOpen
  \bibfield  {author} {\bibinfo {author} {\bibfnamefont {Z.}~\bibnamefont
  {Holmes}}, \bibinfo {author} {\bibfnamefont {K.}~\bibnamefont {Sharma}},
  \bibinfo {author} {\bibfnamefont {M.}~\bibnamefont {Cerezo}},\ and\ \bibinfo
  {author} {\bibfnamefont {P.~J.}\ \bibnamefont {Coles}},\ }\bibfield  {title}
  {\bibinfo {title} {{Connecting Ansatz Expressibility to Gradient Magnitudes
  and Barren Plateaus}},\ }\href {https://doi.org/10.1103/PRXQuantum.3.010313}
  {\bibfield  {journal} {\bibinfo  {journal} {PRX Quantum}\ }\textbf {\bibinfo
  {volume} {3}},\ \bibinfo {pages} {010313} (\bibinfo {year}
  {2022})}\BibitemShut {NoStop}%
\bibitem [{\citenamefont {Arrasmith}\ \emph {et~al.}(2022)\citenamefont
  {Arrasmith}, \citenamefont {Holmes}, \citenamefont {Cerezo},\ and\
  \citenamefont {Coles}}]{Arrasmith2022}%
  \BibitemOpen
  \bibfield  {author} {\bibinfo {author} {\bibfnamefont {A.}~\bibnamefont
  {Arrasmith}}, \bibinfo {author} {\bibfnamefont {Z.}~\bibnamefont {Holmes}},
  \bibinfo {author} {\bibfnamefont {M.}~\bibnamefont {Cerezo}},\ and\ \bibinfo
  {author} {\bibfnamefont {P.~J.}\ \bibnamefont {Coles}},\ }\bibfield  {title}
  {\bibinfo {title} {Equivalence of quantum barren plateaus to cost
  concentration and narrow gorges},\ }\href
  {https://doi.org/10.1088/2058-9565/ac7d06} {\bibfield  {journal} {\bibinfo
  {journal} {Quantum Sci. Technol.}\ }\textbf {\bibinfo {volume} {7}},\
  \bibinfo {pages} {045015} (\bibinfo {year} {2022})}\BibitemShut {NoStop}%
\bibitem [{\citenamefont {Ba\~nuls}\ \emph {et~al.}(2006)\citenamefont
  {Ba\~nuls}, \citenamefont {Or\'us}, \citenamefont {Latorre}, \citenamefont
  {P\'erez},\ and\ \citenamefont {Ruiz-Femen\'{\i}a}}]{BaEtAl06}%
  \BibitemOpen
  \bibfield  {author} {\bibinfo {author} {\bibfnamefont {M.~C.}\ \bibnamefont
  {Ba\~nuls}}, \bibinfo {author} {\bibfnamefont {R.}~\bibnamefont {Or\'us}},
  \bibinfo {author} {\bibfnamefont {J.~I.}\ \bibnamefont {Latorre}}, \bibinfo
  {author} {\bibfnamefont {A.}~\bibnamefont {P\'erez}},\ and\ \bibinfo {author}
  {\bibfnamefont {P.}~\bibnamefont {Ruiz-Femen\'{\i}a}},\ }\bibfield  {title}
  {\bibinfo {title} {Simulation of many-qubit quantum computation with matrix
  product states},\ }\href {https://doi.org/10.1103/PhysRevA.73.022344}
  {\bibfield  {journal} {\bibinfo  {journal} {Phys. Rev. A}\ }\textbf {\bibinfo
  {volume} {73}},\ \bibinfo {pages} {022344} (\bibinfo {year}
  {2006})}\BibitemShut {NoStop}%
\bibitem [{\citenamefont {Bauer}\ \emph {et~al.}(2015)\citenamefont {Bauer},
  \citenamefont {Wang}, \citenamefont {Pižorn},\ and\ \citenamefont
  {Troyer}}]{BaEtAl15}%
  \BibitemOpen
  \bibfield  {author} {\bibinfo {author} {\bibfnamefont {B.}~\bibnamefont
  {Bauer}}, \bibinfo {author} {\bibfnamefont {L.}~\bibnamefont {Wang}},
  \bibinfo {author} {\bibfnamefont {I.}~\bibnamefont {Pižorn}},\ and\ \bibinfo
  {author} {\bibfnamefont {M.}~\bibnamefont {Troyer}},\ }\href@noop {}
  {\bibinfo {title} {Entanglement as a resource in adiabatic quantum
  optimization}} (\bibinfo {year} {2015}),\ \Eprint
  {https://arxiv.org/abs/1501.06914} {arXiv:1501.06914 [cond-mat.dis-nn]}
  \BibitemShut {NoStop}%
\bibitem [{\citenamefont {Lami}\ \emph {et~al.}(2023)\citenamefont {Lami},
  \citenamefont {Torta}, \citenamefont {Santoro},\ and\ \citenamefont
  {Collura}}]{LaEtAl23}%
  \BibitemOpen
  \bibfield  {author} {\bibinfo {author} {\bibfnamefont {G.}~\bibnamefont
  {Lami}}, \bibinfo {author} {\bibfnamefont {P.}~\bibnamefont {Torta}},
  \bibinfo {author} {\bibfnamefont {G.~E.}\ \bibnamefont {Santoro}},\ and\
  \bibinfo {author} {\bibfnamefont {M.}~\bibnamefont {Collura}},\ }\bibfield
  {title} {\bibinfo {title} {Quantum annealing for neural network optimization
  problems: {A} new approach via tensor network simulations},\ }\href
  {https://doi.org/10.21468/SciPostPhys.14.5.117} {\bibfield  {journal}
  {\bibinfo  {journal} {SciPost Phys.}\ }\textbf {\bibinfo {volume} {14}},\
  \bibinfo {pages} {117} (\bibinfo {year} {2023})}\BibitemShut {NoStop}%
\bibitem [{\citenamefont {Granet}\ and\ \citenamefont {Dreyer}(2024)}]{GrDr24}%
  \BibitemOpen
  \bibfield  {author} {\bibinfo {author} {\bibfnamefont {E.}~\bibnamefont
  {Granet}}\ and\ \bibinfo {author} {\bibfnamefont {H.}~\bibnamefont
  {Dreyer}},\ }\href@noop {} {\bibinfo {title} {Benchmarking a heuristic
  {Floquet} adiabatic algorithm for the {Max-Cut} problem}} (\bibinfo {year}
  {2024}),\ \Eprint {https://arxiv.org/abs/2404.16001} {arXiv:2404.16001
  [quant-ph]} \BibitemShut {NoStop}%
\bibitem [{\citenamefont {Oseledets}\ and\ \citenamefont
  {Dolgov}(2012)}]{OsDo12}%
  \BibitemOpen
  \bibfield  {author} {\bibinfo {author} {\bibfnamefont {I.~V.}\ \bibnamefont
  {Oseledets}}\ and\ \bibinfo {author} {\bibfnamefont {S.~V.}\ \bibnamefont
  {Dolgov}},\ }\bibfield  {title} {\bibinfo {title} {{Solution of Linear
  Systems and Matrix Inversion in the TT-Format}},\ }\href
  {https://doi.org/10.1137/110833142} {\bibfield  {journal} {\bibinfo
  {journal} {SIAM J. Sci. Comput.}\ }\textbf {\bibinfo {volume} {34}},\
  \bibinfo {pages} {A2718} (\bibinfo {year} {2012})}\BibitemShut {NoStop}%
\bibitem [{\citenamefont {Lubasch}\ \emph {et~al.}(2018)\citenamefont
  {Lubasch}, \citenamefont {Moinier},\ and\ \citenamefont {Jaksch}}]{LuMoJa18}%
  \BibitemOpen
  \bibfield  {author} {\bibinfo {author} {\bibfnamefont {M.}~\bibnamefont
  {Lubasch}}, \bibinfo {author} {\bibfnamefont {P.}~\bibnamefont {Moinier}},\
  and\ \bibinfo {author} {\bibfnamefont {D.}~\bibnamefont {Jaksch}},\
  }\bibfield  {title} {\bibinfo {title} {Multigrid renormalization},\ }\href
  {https://doi.org/https://doi.org/10.1016/j.jcp.2018.06.065} {\bibfield
  {journal} {\bibinfo  {journal} {J. Comp. Phys.}\ }\textbf {\bibinfo {volume}
  {372}},\ \bibinfo {pages} {587} (\bibinfo {year} {2018})}\BibitemShut
  {NoStop}%
\bibitem [{\citenamefont {Zaletel}\ \emph {et~al.}(2015)\citenamefont
  {Zaletel}, \citenamefont {Mong}, \citenamefont {Karrasch}, \citenamefont
  {Moore},\ and\ \citenamefont {Pollmann}}]{ZaletelMPO}%
  \BibitemOpen
  \bibfield  {author} {\bibinfo {author} {\bibfnamefont {M.~P.}\ \bibnamefont
  {Zaletel}}, \bibinfo {author} {\bibfnamefont {R.~S.~K.}\ \bibnamefont
  {Mong}}, \bibinfo {author} {\bibfnamefont {C.}~\bibnamefont {Karrasch}},
  \bibinfo {author} {\bibfnamefont {J.~E.}\ \bibnamefont {Moore}},\ and\
  \bibinfo {author} {\bibfnamefont {F.}~\bibnamefont {Pollmann}},\ }\bibfield
  {title} {\bibinfo {title} {Time-evolving a matrix product state with
  long-ranged interactions},\ }\href
  {https://doi.org/10.1103/PhysRevB.91.165112} {\bibfield  {journal} {\bibinfo
  {journal} {Phys. Rev. B}\ }\textbf {\bibinfo {volume} {91}},\ \bibinfo
  {pages} {165112} (\bibinfo {year} {2015})}\BibitemShut {NoStop}%
\bibitem [{\citenamefont {Van~Damme}\ \emph {et~al.}(2023)\citenamefont
  {Van~Damme}, \citenamefont {Haegeman}, \citenamefont {McCulloch},\ and\
  \citenamefont {Vanderstraeten}}]{VaEtAl23}%
  \BibitemOpen
  \bibfield  {author} {\bibinfo {author} {\bibfnamefont {M.}~\bibnamefont
  {Van~Damme}}, \bibinfo {author} {\bibfnamefont {J.}~\bibnamefont {Haegeman}},
  \bibinfo {author} {\bibfnamefont {I.}~\bibnamefont {McCulloch}},\ and\
  \bibinfo {author} {\bibfnamefont {L.}~\bibnamefont {Vanderstraeten}},\
  }\href@noop {} {\bibinfo {title} {Efficient higher-order matrix product
  operators for time evolution}} (\bibinfo {year} {2023}),\ \Eprint
  {https://arxiv.org/abs/2302.14181} {arXiv:2302.14181 [cond-mat.str-el]}
  \BibitemShut {NoStop}%
\bibitem [{\citenamefont {Takahashi}\ and\ \citenamefont {del
  Campo}(2024)}]{TaDe23}%
  \BibitemOpen
  \bibfield  {author} {\bibinfo {author} {\bibfnamefont {K.}~\bibnamefont
  {Takahashi}}\ and\ \bibinfo {author} {\bibfnamefont {A.}~\bibnamefont {del
  Campo}},\ }\bibfield  {title} {\bibinfo {title} {{Shortcuts to Adiabaticity
  in Krylov Space}},\ }\href {https://doi.org/10.1103/PhysRevX.14.011032}
  {\bibfield  {journal} {\bibinfo  {journal} {Phys. Rev. X}\ }\textbf {\bibinfo
  {volume} {14}},\ \bibinfo {pages} {011032} (\bibinfo {year}
  {2024})}\BibitemShut {NoStop}%
\bibitem [{\citenamefont {Feist}\ and\ \citenamefont
  {contributors}(2021)}]{QuantumAlgebra.jl}%
  \BibitemOpen
  \bibfield  {author} {\bibinfo {author} {\bibfnamefont {J.}~\bibnamefont
  {Feist}}\ and\ \bibinfo {author} {\bibnamefont {contributors}},\ }\href
  {https://doi.org/10.5281/zenodo.3525845} {\bibinfo {title}
  {Quantumalgebra.jl}} (\bibinfo {year} {2021})\BibitemShut {NoStop}%
\bibitem [{\citenamefont {Fishman}\ \emph
  {et~al.}(2022{\natexlab{a}})\citenamefont {Fishman}, \citenamefont {White},\
  and\ \citenamefont {Stoudenmire}}]{ITensor}%
  \BibitemOpen
  \bibfield  {author} {\bibinfo {author} {\bibfnamefont {M.}~\bibnamefont
  {Fishman}}, \bibinfo {author} {\bibfnamefont {S.~R.}\ \bibnamefont {White}},\
  and\ \bibinfo {author} {\bibfnamefont {E.~M.}\ \bibnamefont {Stoudenmire}},\
  }\bibfield  {title} {\bibinfo {title} {{The ITensor Software Library for
  Tensor Network Calculations}},\ }\href
  {https://doi.org/10.21468/SciPostPhysCodeb.4} {\bibfield  {journal} {\bibinfo
   {journal} {SciPost Phys. Codebases}\ ,\ \bibinfo {pages} {4}} (\bibinfo
  {year} {2022}{\natexlab{a}})}\BibitemShut {NoStop}%
\bibitem [{\citenamefont {Fishman}\ \emph
  {et~al.}(2022{\natexlab{b}})\citenamefont {Fishman}, \citenamefont {White},\
  and\ \citenamefont {Stoudenmire}}]{ITensor-r0.3}%
  \BibitemOpen
  \bibfield  {author} {\bibinfo {author} {\bibfnamefont {M.}~\bibnamefont
  {Fishman}}, \bibinfo {author} {\bibfnamefont {S.~R.}\ \bibnamefont {White}},\
  and\ \bibinfo {author} {\bibfnamefont {E.~M.}\ \bibnamefont {Stoudenmire}},\
  }\bibfield  {title} {\bibinfo {title} {{Codebase release 0.3 for ITensor}},\
  }\href {https://doi.org/10.21468/SciPostPhysCodeb.4-r0.3} {\bibfield
  {journal} {\bibinfo  {journal} {SciPost Phys. Codebases}\ ,\ \bibinfo {pages}
  {4}} (\bibinfo {year} {2022}{\natexlab{b}})}\BibitemShut {NoStop}%
\bibitem [{\citenamefont {Schollw\"ock}(2005)}]{Sc05}%
  \BibitemOpen
  \bibfield  {author} {\bibinfo {author} {\bibfnamefont {U.}~\bibnamefont
  {Schollw\"ock}},\ }\bibfield  {title} {\bibinfo {title} {The density-matrix
  renormalization group},\ }\href {https://doi.org/10.1103/RevModPhys.77.259}
  {\bibfield  {journal} {\bibinfo  {journal} {Rev. Mod. Phys.}\ }\textbf
  {\bibinfo {volume} {77}},\ \bibinfo {pages} {259} (\bibinfo {year}
  {2005})}\BibitemShut {NoStop}%
\bibitem [{\citenamefont {Hatano}\ and\ \citenamefont {Suzuki}(2005)}]{HaSu05}%
  \BibitemOpen
  \bibfield  {author} {\bibinfo {author} {\bibfnamefont {N.}~\bibnamefont
  {Hatano}}\ and\ \bibinfo {author} {\bibfnamefont {M.}~\bibnamefont
  {Suzuki}},\ }\bibinfo {title} {Finding exponential product formulas of higher
  orders},\ in\ \href {https://doi.org/https://doi.org/10.1007/11526216_2}
  {\emph {\bibinfo {booktitle} {Quantum Annealing and Other Optimization
  Methods}}}\ (\bibinfo  {publisher} {Springer Berlin Heidelberg},\ \bibinfo
  {address} {Berlin, Heidelberg},\ \bibinfo {year} {2005})\ pp.\ \bibinfo
  {pages} {37--68}\BibitemShut {NoStop}%
\bibitem [{\citenamefont {Lubasch}\ \emph {et~al.}(2011)\citenamefont
  {Lubasch}, \citenamefont {Murg}, \citenamefont {Schneider}, \citenamefont
  {Cirac},\ and\ \citenamefont {Ba\~nuls}}]{LuEtAl11}%
  \BibitemOpen
  \bibfield  {author} {\bibinfo {author} {\bibfnamefont {M.}~\bibnamefont
  {Lubasch}}, \bibinfo {author} {\bibfnamefont {V.}~\bibnamefont {Murg}},
  \bibinfo {author} {\bibfnamefont {U.}~\bibnamefont {Schneider}}, \bibinfo
  {author} {\bibfnamefont {J.~I.}\ \bibnamefont {Cirac}},\ and\ \bibinfo
  {author} {\bibfnamefont {M.~C.}\ \bibnamefont {Ba\~nuls}},\ }\bibfield
  {title} {\bibinfo {title} {{Adiabatic Preparation of a Heisenberg
  Antiferromagnet Using an Optical Superlattice}},\ }\href
  {https://doi.org/10.1103/PhysRevLett.107.165301} {\bibfield  {journal}
  {\bibinfo  {journal} {Phys. Rev. Lett.}\ }\textbf {\bibinfo {volume} {107}},\
  \bibinfo {pages} {165301} (\bibinfo {year} {2011})}\BibitemShut {NoStop}%
\bibitem [{\citenamefont {Ran}(2020)}]{Ra20}%
  \BibitemOpen
  \bibfield  {author} {\bibinfo {author} {\bibfnamefont {S.-J.}\ \bibnamefont
  {Ran}},\ }\bibfield  {title} {\bibinfo {title} {Encoding of matrix product
  states into quantum circuits of one- and two-qubit gates},\ }\href
  {https://doi.org/10.1103/PhysRevA.101.032310} {\bibfield  {journal} {\bibinfo
   {journal} {Phys. Rev. A}\ }\textbf {\bibinfo {volume} {101}},\ \bibinfo
  {pages} {032310} (\bibinfo {year} {2020})}\BibitemShut {NoStop}%
\bibitem [{\citenamefont {Shende}\ \emph {et~al.}(2004)\citenamefont {Shende},
  \citenamefont {Markov},\ and\ \citenamefont {Bullock}}]{ShMaBu04}%
  \BibitemOpen
  \bibfield  {author} {\bibinfo {author} {\bibfnamefont {V.~V.}\ \bibnamefont
  {Shende}}, \bibinfo {author} {\bibfnamefont {I.~L.}\ \bibnamefont {Markov}},\
  and\ \bibinfo {author} {\bibfnamefont {S.~S.}\ \bibnamefont {Bullock}},\
  }\bibfield  {title} {\bibinfo {title} {Minimal universal two-qubit
  controlled-$not$-based circuits},\ }\href
  {https://doi.org/10.1103/PhysRevA.69.062321} {\bibfield  {journal} {\bibinfo
  {journal} {Phys. Rev. A}\ }\textbf {\bibinfo {volume} {69}},\ \bibinfo
  {pages} {062321} (\bibinfo {year} {2004})}\BibitemShut {NoStop}%
\bibitem [{\citenamefont {Sch\"on}\ \emph {et~al.}(2005)\citenamefont
  {Sch\"on}, \citenamefont {Solano}, \citenamefont {Verstraete}, \citenamefont
  {Cirac},\ and\ \citenamefont {Wolf}}]{ScEtAl05}%
  \BibitemOpen
  \bibfield  {author} {\bibinfo {author} {\bibfnamefont {C.}~\bibnamefont
  {Sch\"on}}, \bibinfo {author} {\bibfnamefont {E.}~\bibnamefont {Solano}},
  \bibinfo {author} {\bibfnamefont {F.}~\bibnamefont {Verstraete}}, \bibinfo
  {author} {\bibfnamefont {J.~I.}\ \bibnamefont {Cirac}},\ and\ \bibinfo
  {author} {\bibfnamefont {M.~M.}\ \bibnamefont {Wolf}},\ }\bibfield  {title}
  {\bibinfo {title} {{Sequential Generation of Entangled Multiqubit States}},\
  }\href {https://doi.org/10.1103/PhysRevLett.95.110503} {\bibfield  {journal}
  {\bibinfo  {journal} {Phys. Rev. Lett.}\ }\textbf {\bibinfo {volume} {95}},\
  \bibinfo {pages} {110503} (\bibinfo {year} {2005})}\BibitemShut {NoStop}%
\bibitem [{\citenamefont {Sch\"on}\ \emph {et~al.}(2007)\citenamefont
  {Sch\"on}, \citenamefont {Hammerer}, \citenamefont {Wolf}, \citenamefont
  {Cirac},\ and\ \citenamefont {Solano}}]{ScEtAl07}%
  \BibitemOpen
  \bibfield  {author} {\bibinfo {author} {\bibfnamefont {C.}~\bibnamefont
  {Sch\"on}}, \bibinfo {author} {\bibfnamefont {K.}~\bibnamefont {Hammerer}},
  \bibinfo {author} {\bibfnamefont {M.~M.}\ \bibnamefont {Wolf}}, \bibinfo
  {author} {\bibfnamefont {J.~I.}\ \bibnamefont {Cirac}},\ and\ \bibinfo
  {author} {\bibfnamefont {E.}~\bibnamefont {Solano}},\ }\bibfield  {title}
  {\bibinfo {title} {Sequential generation of matrix-product states in cavity
  {QED}},\ }\href {https://doi.org/10.1103/PhysRevA.75.032311} {\bibfield
  {journal} {\bibinfo  {journal} {Phys. Rev. A}\ }\textbf {\bibinfo {volume}
  {75}},\ \bibinfo {pages} {032311} (\bibinfo {year} {2007})}\BibitemShut
  {NoStop}%
\bibitem [{\citenamefont {Lubasch}\ \emph {et~al.}(2020)\citenamefont
  {Lubasch}, \citenamefont {Joo}, \citenamefont {Moinier}, \citenamefont
  {Kiffner},\ and\ \citenamefont {Jaksch}}]{LuEtAl20}%
  \BibitemOpen
  \bibfield  {author} {\bibinfo {author} {\bibfnamefont {M.}~\bibnamefont
  {Lubasch}}, \bibinfo {author} {\bibfnamefont {J.}~\bibnamefont {Joo}},
  \bibinfo {author} {\bibfnamefont {P.}~\bibnamefont {Moinier}}, \bibinfo
  {author} {\bibfnamefont {M.}~\bibnamefont {Kiffner}},\ and\ \bibinfo {author}
  {\bibfnamefont {D.}~\bibnamefont {Jaksch}},\ }\bibfield  {title} {\bibinfo
  {title} {Variational quantum algorithms for nonlinear problems},\ }\href
  {https://doi.org/10.1103/PhysRevA.101.010301} {\bibfield  {journal} {\bibinfo
   {journal} {Phys. Rev. A}\ }\textbf {\bibinfo {volume} {101}},\ \bibinfo
  {pages} {010301} (\bibinfo {year} {2020})}\BibitemShut {NoStop}%
\bibitem [{\citenamefont {Akhalwaya}\ \emph {et~al.}(2023)\citenamefont
  {Akhalwaya}, \citenamefont {Connolly}, \citenamefont {Guichard},
  \citenamefont {Herbert}, \citenamefont {Kargi}, \citenamefont {Krajenbrink},
  \citenamefont {Lubasch}, \citenamefont {Keever}, \citenamefont {Sorci},
  \citenamefont {Spranger},\ and\ \citenamefont {Williams}}]{AkEtAl23}%
  \BibitemOpen
  \bibfield  {author} {\bibinfo {author} {\bibfnamefont {I.~Y.}\ \bibnamefont
  {Akhalwaya}}, \bibinfo {author} {\bibfnamefont {A.}~\bibnamefont {Connolly}},
  \bibinfo {author} {\bibfnamefont {R.}~\bibnamefont {Guichard}}, \bibinfo
  {author} {\bibfnamefont {S.}~\bibnamefont {Herbert}}, \bibinfo {author}
  {\bibfnamefont {C.}~\bibnamefont {Kargi}}, \bibinfo {author} {\bibfnamefont
  {A.}~\bibnamefont {Krajenbrink}}, \bibinfo {author} {\bibfnamefont
  {M.}~\bibnamefont {Lubasch}}, \bibinfo {author} {\bibfnamefont {C.~M.}\
  \bibnamefont {Keever}}, \bibinfo {author} {\bibfnamefont {J.}~\bibnamefont
  {Sorci}}, \bibinfo {author} {\bibfnamefont {M.}~\bibnamefont {Spranger}},\
  and\ \bibinfo {author} {\bibfnamefont {I.}~\bibnamefont {Williams}},\
  }\href@noop {} {\bibinfo {title} {{A Modular Engine for Quantum Monte Carlo
  Integration}}} (\bibinfo {year} {2023}),\ \Eprint
  {https://arxiv.org/abs/2308.06081} {arXiv:2308.06081 [quant-ph]} \BibitemShut
  {NoStop}%
\bibitem [{\citenamefont {Malz}\ \emph {et~al.}(2024)\citenamefont {Malz},
  \citenamefont {Styliaris}, \citenamefont {Wei},\ and\ \citenamefont
  {Cirac}}]{MaStWeCi24}%
  \BibitemOpen
  \bibfield  {author} {\bibinfo {author} {\bibfnamefont {D.}~\bibnamefont
  {Malz}}, \bibinfo {author} {\bibfnamefont {G.}~\bibnamefont {Styliaris}},
  \bibinfo {author} {\bibfnamefont {Z.-Y.}\ \bibnamefont {Wei}},\ and\ \bibinfo
  {author} {\bibfnamefont {J.~I.}\ \bibnamefont {Cirac}},\ }\bibfield  {title}
  {\bibinfo {title} {{Preparation of Matrix Product States with Log-Depth
  Quantum Circuits}},\ }\href {https://doi.org/10.1103/PhysRevLett.132.040404}
  {\bibfield  {journal} {\bibinfo  {journal} {Phys. Rev. Lett.}\ }\textbf
  {\bibinfo {volume} {132}},\ \bibinfo {pages} {040404} (\bibinfo {year}
  {2024})}\BibitemShut {NoStop}%
\bibitem [{\citenamefont {Stoudenmire}\ and\ \citenamefont
  {White}(2012)}]{StWh12}%
  \BibitemOpen
  \bibfield  {author} {\bibinfo {author} {\bibfnamefont {E.~M.}\ \bibnamefont
  {Stoudenmire}}\ and\ \bibinfo {author} {\bibfnamefont {S.~R.}\ \bibnamefont
  {White}},\ }\bibfield  {title} {\bibinfo {title} {{Studying Two-Dimensional
  Systems with the Density Matrix Renormalization Group}},\ }\href
  {https://doi.org/10.1146/annurev-conmatphys-020911-125018} {\bibfield
  {journal} {\bibinfo  {journal} {Annu. Rev. Condens. Matter Phys.}\ }\textbf
  {\bibinfo {volume} {3}},\ \bibinfo {pages} {111} (\bibinfo {year}
  {2012})}\BibitemShut {NoStop}%
\bibitem [{\citenamefont {Dupont}\ \emph
  {et~al.}(2022{\natexlab{a}})\citenamefont {Dupont}, \citenamefont {Didier},
  \citenamefont {Hodson}, \citenamefont {Moore},\ and\ \citenamefont
  {Reagor}}]{DuEtAl22a}%
  \BibitemOpen
  \bibfield  {author} {\bibinfo {author} {\bibfnamefont {M.}~\bibnamefont
  {Dupont}}, \bibinfo {author} {\bibfnamefont {N.}~\bibnamefont {Didier}},
  \bibinfo {author} {\bibfnamefont {M.~J.}\ \bibnamefont {Hodson}}, \bibinfo
  {author} {\bibfnamefont {J.~E.}\ \bibnamefont {Moore}},\ and\ \bibinfo
  {author} {\bibfnamefont {M.~J.}\ \bibnamefont {Reagor}},\ }\bibfield  {title}
  {\bibinfo {title} {Entanglement perspective on the quantum approximate
  optimization algorithm},\ }\href
  {https://doi.org/10.1103/PhysRevA.106.022423} {\bibfield  {journal} {\bibinfo
   {journal} {Phys. Rev. A}\ }\textbf {\bibinfo {volume} {106}},\ \bibinfo
  {pages} {022423} (\bibinfo {year} {2022}{\natexlab{a}})}\BibitemShut
  {NoStop}%
\bibitem [{\citenamefont {Dupont}\ \emph
  {et~al.}(2022{\natexlab{b}})\citenamefont {Dupont}, \citenamefont {Didier},
  \citenamefont {Hodson}, \citenamefont {Moore},\ and\ \citenamefont
  {Reagor}}]{DuEtAl22b}%
  \BibitemOpen
  \bibfield  {author} {\bibinfo {author} {\bibfnamefont {M.}~\bibnamefont
  {Dupont}}, \bibinfo {author} {\bibfnamefont {N.}~\bibnamefont {Didier}},
  \bibinfo {author} {\bibfnamefont {M.~J.}\ \bibnamefont {Hodson}}, \bibinfo
  {author} {\bibfnamefont {J.~E.}\ \bibnamefont {Moore}},\ and\ \bibinfo
  {author} {\bibfnamefont {M.~J.}\ \bibnamefont {Reagor}},\ }\bibfield  {title}
  {\bibinfo {title} {{Calibrating the Classical Hardness of the Quantum
  Approximate Optimization Algorithm}},\ }\href
  {https://doi.org/10.1103/PRXQuantum.3.040339} {\bibfield  {journal} {\bibinfo
   {journal} {PRX Quantum}\ }\textbf {\bibinfo {volume} {3}},\ \bibinfo {pages}
  {040339} (\bibinfo {year} {2022}{\natexlab{b}})}\BibitemShut {NoStop}%
\bibitem [{\citenamefont {Verstraete}\ and\ \citenamefont
  {Cirac}(2004)}]{VeCi04}%
  \BibitemOpen
  \bibfield  {author} {\bibinfo {author} {\bibfnamefont {F.}~\bibnamefont
  {Verstraete}}\ and\ \bibinfo {author} {\bibfnamefont {J.~I.}\ \bibnamefont
  {Cirac}},\ }\href@noop {} {\bibinfo {title} {Renormalization algorithms for
  quantum many-body systems in two and higher dimensions}} (\bibinfo {year}
  {2004}),\ \Eprint {https://arxiv.org/abs/cond-mat/0407066}
  {arXiv:cond-mat/0407066 [cond-mat.str-el]} \BibitemShut {NoStop}%
\bibitem [{\citenamefont {Lubasch}\ \emph
  {et~al.}(2014{\natexlab{a}})\citenamefont {Lubasch}, \citenamefont {Cirac},\
  and\ \citenamefont {Bañuls}}]{LuCiBa14a}%
  \BibitemOpen
  \bibfield  {author} {\bibinfo {author} {\bibfnamefont {M.}~\bibnamefont
  {Lubasch}}, \bibinfo {author} {\bibfnamefont {J.~I.}\ \bibnamefont {Cirac}},\
  and\ \bibinfo {author} {\bibfnamefont {M.~C.}\ \bibnamefont {Bañuls}},\
  }\bibfield  {title} {\bibinfo {title} {Unifying projected entangled pair
  state contractions},\ }\href {https://doi.org/10.1088/1367-2630/16/3/033014}
  {\bibfield  {journal} {\bibinfo  {journal} {New J. Phys.}\ }\textbf {\bibinfo
  {volume} {16}},\ \bibinfo {pages} {033014} (\bibinfo {year}
  {2014}{\natexlab{a}})}\BibitemShut {NoStop}%
\bibitem [{\citenamefont {Lubasch}\ \emph
  {et~al.}(2014{\natexlab{b}})\citenamefont {Lubasch}, \citenamefont {Cirac},\
  and\ \citenamefont {Ba\~nuls}}]{LuCiBa14b}%
  \BibitemOpen
  \bibfield  {author} {\bibinfo {author} {\bibfnamefont {M.}~\bibnamefont
  {Lubasch}}, \bibinfo {author} {\bibfnamefont {J.~I.}\ \bibnamefont {Cirac}},\
  and\ \bibinfo {author} {\bibfnamefont {M.~C.}\ \bibnamefont {Ba\~nuls}},\
  }\bibfield  {title} {\bibinfo {title} {Algorithms for finite projected
  entangled pair states},\ }\href {https://doi.org/10.1103/PhysRevB.90.064425}
  {\bibfield  {journal} {\bibinfo  {journal} {Phys. Rev. B}\ }\textbf {\bibinfo
  {volume} {90}},\ \bibinfo {pages} {064425} (\bibinfo {year}
  {2014}{\natexlab{b}})}\BibitemShut {NoStop}%
\bibitem [{\citenamefont {Czarnik}\ \emph {et~al.}(2019)\citenamefont
  {Czarnik}, \citenamefont {Dziarmaga},\ and\ \citenamefont
  {Corboz}}]{CzDzCo19}%
  \BibitemOpen
  \bibfield  {author} {\bibinfo {author} {\bibfnamefont {P.}~\bibnamefont
  {Czarnik}}, \bibinfo {author} {\bibfnamefont {J.}~\bibnamefont {Dziarmaga}},\
  and\ \bibinfo {author} {\bibfnamefont {P.}~\bibnamefont {Corboz}},\
  }\bibfield  {title} {\bibinfo {title} {Time evolution of an infinite
  projected entangled pair state: {An} efficient algorithm},\ }\href
  {https://doi.org/10.1103/PhysRevB.99.035115} {\bibfield  {journal} {\bibinfo
  {journal} {Phys. Rev. B}\ }\textbf {\bibinfo {volume} {99}},\ \bibinfo
  {pages} {035115} (\bibinfo {year} {2019})}\BibitemShut {NoStop}%
\bibitem [{\citenamefont {Mc~Keever}\ and\ \citenamefont
  {Szyma\ifmmode~\acute{n}\else \'{n}\fi{}ska}(2021)}]{McSz21}%
  \BibitemOpen
  \bibfield  {author} {\bibinfo {author} {\bibfnamefont {C.}~\bibnamefont
  {Mc~Keever}}\ and\ \bibinfo {author} {\bibfnamefont {M.~H.}\ \bibnamefont
  {Szyma\ifmmode~\acute{n}\else \'{n}\fi{}ska}},\ }\bibfield  {title} {\bibinfo
  {title} {{Stable iPEPO Tensor-Network Algorithm for Dynamics of
  Two-Dimensional Open Quantum Lattice Models}},\ }\href
  {https://doi.org/10.1103/PhysRevX.11.021035} {\bibfield  {journal} {\bibinfo
  {journal} {Phys. Rev. X}\ }\textbf {\bibinfo {volume} {11}},\ \bibinfo
  {pages} {021035} (\bibinfo {year} {2021})}\BibitemShut {NoStop}%
\bibitem [{\citenamefont {Lucas}(2014)}]{Lu14}%
  \BibitemOpen
  \bibfield  {author} {\bibinfo {author} {\bibfnamefont {A.}~\bibnamefont
  {Lucas}},\ }\bibfield  {title} {\bibinfo {title} {Ising formulations of many
  {NP} problems},\ }\href
  {https://www.frontiersin.org/articles/10.3389/fphy.2014.00005} {\bibfield
  {journal} {\bibinfo  {journal} {Front. Phys.}\ }\textbf {\bibinfo {volume}
  {2}} (\bibinfo {year} {2014})}\BibitemShut {NoStop}%
\bibitem [{\citenamefont {Glover}\ \emph {et~al.}(2022)\citenamefont {Glover},
  \citenamefont {Kochenberger}, \citenamefont {Hennig},\ and\ \citenamefont
  {Du}}]{GlEtAl22}%
  \BibitemOpen
  \bibfield  {author} {\bibinfo {author} {\bibfnamefont {F.}~\bibnamefont
  {Glover}}, \bibinfo {author} {\bibfnamefont {G.}~\bibnamefont
  {Kochenberger}}, \bibinfo {author} {\bibfnamefont {R.}~\bibnamefont
  {Hennig}},\ and\ \bibinfo {author} {\bibfnamefont {Y.}~\bibnamefont {Du}},\
  }\bibfield  {title} {\bibinfo {title} {Quantum bridge analytics {I}: a
  tutorial on formulating and using {QUBO} models},\ }\href
  {https://doi.org/10.1007/s10479-022-04634-2} {\bibfield  {journal} {\bibinfo
  {journal} {Ann. Oper. Res.}\ }\textbf {\bibinfo {volume} {314}},\ \bibinfo
  {pages} {141} (\bibinfo {year} {2022})}\BibitemShut {NoStop}%
\bibitem [{\citenamefont {Haghshenas}\ \emph {et~al.}(2022)\citenamefont
  {Haghshenas}, \citenamefont {Gray}, \citenamefont {Potter},\ and\
  \citenamefont {Chan}}]{HaEtAl22}%
  \BibitemOpen
  \bibfield  {author} {\bibinfo {author} {\bibfnamefont {R.}~\bibnamefont
  {Haghshenas}}, \bibinfo {author} {\bibfnamefont {J.}~\bibnamefont {Gray}},
  \bibinfo {author} {\bibfnamefont {A.~C.}\ \bibnamefont {Potter}},\ and\
  \bibinfo {author} {\bibfnamefont {G.~K.-L.}\ \bibnamefont {Chan}},\
  }\bibfield  {title} {\bibinfo {title} {{Variational Power of Quantum Circuit
  Tensor Networks}},\ }\href {https://doi.org/10.1103/PhysRevX.12.011047}
  {\bibfield  {journal} {\bibinfo  {journal} {Phys. Rev. X}\ }\textbf {\bibinfo
  {volume} {12}},\ \bibinfo {pages} {011047} (\bibinfo {year}
  {2022})}\BibitemShut {NoStop}%
\bibitem [{\citenamefont {Cervero~Mart{\'{i}}n}\ \emph
  {et~al.}(2023)\citenamefont {Cervero~Mart{\'{i}}n}, \citenamefont
  {Plekhanov},\ and\ \citenamefont {Lubasch}}]{CePlLu23}%
  \BibitemOpen
  \bibfield  {author} {\bibinfo {author} {\bibfnamefont {E.}~\bibnamefont
  {Cervero~Mart{\'{i}}n}}, \bibinfo {author} {\bibfnamefont {K.}~\bibnamefont
  {Plekhanov}},\ and\ \bibinfo {author} {\bibfnamefont {M.}~\bibnamefont
  {Lubasch}},\ }\bibfield  {title} {\bibinfo {title} {Barren plateaus in
  quantum tensor network optimization},\ }\href
  {https://doi.org/10.22331/q-2023-04-13-974} {\bibfield  {journal} {\bibinfo
  {journal} {Quantum}\ }\textbf {\bibinfo {volume} {7}},\ \bibinfo {pages}
  {974} (\bibinfo {year} {2023})}\BibitemShut {NoStop}%
\bibitem [{\citenamefont {Haghshenas}\ \emph {et~al.}(2023)\citenamefont
  {Haghshenas}, \citenamefont {Chertkov}, \citenamefont {DeCross},
  \citenamefont {Gatterman}, \citenamefont {Gerber}, \citenamefont {Gilmore},
  \citenamefont {Gresh}, \citenamefont {Hewitt}, \citenamefont {Horst},
  \citenamefont {Matheny}, \citenamefont {Mengle}, \citenamefont {Neyenhuis},
  \citenamefont {Hayes},\ and\ \citenamefont {Foss-Feig}}]{HaEtAl23}%
  \BibitemOpen
  \bibfield  {author} {\bibinfo {author} {\bibfnamefont {R.}~\bibnamefont
  {Haghshenas}}, \bibinfo {author} {\bibfnamefont {E.}~\bibnamefont
  {Chertkov}}, \bibinfo {author} {\bibfnamefont {M.}~\bibnamefont {DeCross}},
  \bibinfo {author} {\bibfnamefont {T.~M.}\ \bibnamefont {Gatterman}}, \bibinfo
  {author} {\bibfnamefont {J.~A.}\ \bibnamefont {Gerber}}, \bibinfo {author}
  {\bibfnamefont {K.}~\bibnamefont {Gilmore}}, \bibinfo {author} {\bibfnamefont
  {D.}~\bibnamefont {Gresh}}, \bibinfo {author} {\bibfnamefont
  {N.}~\bibnamefont {Hewitt}}, \bibinfo {author} {\bibfnamefont {C.~V.}\
  \bibnamefont {Horst}}, \bibinfo {author} {\bibfnamefont {M.}~\bibnamefont
  {Matheny}}, \bibinfo {author} {\bibfnamefont {T.}~\bibnamefont {Mengle}},
  \bibinfo {author} {\bibfnamefont {B.}~\bibnamefont {Neyenhuis}}, \bibinfo
  {author} {\bibfnamefont {D.}~\bibnamefont {Hayes}},\ and\ \bibinfo {author}
  {\bibfnamefont {M.}~\bibnamefont {Foss-Feig}},\ }\href@noop {} {\bibinfo
  {title} {Probing critical states of matter on a digital quantum computer}}
  (\bibinfo {year} {2023}),\ \Eprint {https://arxiv.org/abs/2305.01650}
  {arXiv:2305.01650 [quant-ph]} \BibitemShut {NoStop}%
\bibitem [{\citenamefont {Herrman}\ \emph {et~al.}(2022)\citenamefont
  {Herrman}, \citenamefont {Lotshaw}, \citenamefont {Ostrowski}, \citenamefont
  {Humble},\ and\ \citenamefont {Siopsis}}]{HeEtAl22}%
  \BibitemOpen
  \bibfield  {author} {\bibinfo {author} {\bibfnamefont {R.}~\bibnamefont
  {Herrman}}, \bibinfo {author} {\bibfnamefont {P.~C.}\ \bibnamefont
  {Lotshaw}}, \bibinfo {author} {\bibfnamefont {J.}~\bibnamefont {Ostrowski}},
  \bibinfo {author} {\bibfnamefont {T.~S.}\ \bibnamefont {Humble}},\ and\
  \bibinfo {author} {\bibfnamefont {G.}~\bibnamefont {Siopsis}},\ }\bibfield
  {title} {\bibinfo {title} {Multi-angle quantum approximate optimization
  algorithm},\ }\href {https://doi.org/10.1038/s41598-022-10555-8} {\bibfield
  {journal} {\bibinfo  {journal} {Sci. Rep.}\ }\textbf {\bibinfo {volume}
  {12}},\ \bibinfo {pages} {6781} (\bibinfo {year} {2022})}\BibitemShut
  {NoStop}%
\bibitem [{\citenamefont {Anders}\ \emph {et~al.}(2006)\citenamefont {Anders},
  \citenamefont {Plenio}, \citenamefont {D\"ur}, \citenamefont {Verstraete},\
  and\ \citenamefont {Briegel}}]{AnEtAl06}%
  \BibitemOpen
  \bibfield  {author} {\bibinfo {author} {\bibfnamefont {S.}~\bibnamefont
  {Anders}}, \bibinfo {author} {\bibfnamefont {M.~B.}\ \bibnamefont {Plenio}},
  \bibinfo {author} {\bibfnamefont {W.}~\bibnamefont {D\"ur}}, \bibinfo
  {author} {\bibfnamefont {F.}~\bibnamefont {Verstraete}},\ and\ \bibinfo
  {author} {\bibfnamefont {H.-J.}\ \bibnamefont {Briegel}},\ }\bibfield
  {title} {\bibinfo {title} {{Ground-State Approximation for Strongly
  Interacting Spin Systems in Arbitrary Spatial Dimension}},\ }\href
  {https://doi.org/10.1103/PhysRevLett.97.107206} {\bibfield  {journal}
  {\bibinfo  {journal} {Phys. Rev. Lett.}\ }\textbf {\bibinfo {volume} {97}},\
  \bibinfo {pages} {107206} (\bibinfo {year} {2006})}\BibitemShut {NoStop}%
\bibitem [{\citenamefont {Anders}\ \emph {et~al.}(2007)\citenamefont {Anders},
  \citenamefont {Briegel},\ and\ \citenamefont {Dür}}]{AnBrDu07}%
  \BibitemOpen
  \bibfield  {author} {\bibinfo {author} {\bibfnamefont {S.}~\bibnamefont
  {Anders}}, \bibinfo {author} {\bibfnamefont {H.~J.}\ \bibnamefont
  {Briegel}},\ and\ \bibinfo {author} {\bibfnamefont {W.}~\bibnamefont
  {Dür}},\ }\bibfield  {title} {\bibinfo {title} {A variational method based
  on weighted graph states},\ }\href
  {https://doi.org/10.1088/1367-2630/9/10/361} {\bibfield  {journal} {\bibinfo
  {journal} {New J. Phys.}\ }\textbf {\bibinfo {volume} {9}},\ \bibinfo {pages}
  {361} (\bibinfo {year} {2007})}\BibitemShut {NoStop}%
\bibitem [{\citenamefont {Hartmann}\ \emph {et~al.}(2007)\citenamefont
  {Hartmann}, \citenamefont {Calsamiglia}, \citenamefont {Dür},\ and\
  \citenamefont {Briegel}}]{HaEtAl07}%
  \BibitemOpen
  \bibfield  {author} {\bibinfo {author} {\bibfnamefont {L.}~\bibnamefont
  {Hartmann}}, \bibinfo {author} {\bibfnamefont {J.}~\bibnamefont
  {Calsamiglia}}, \bibinfo {author} {\bibfnamefont {W.}~\bibnamefont {Dür}},\
  and\ \bibinfo {author} {\bibfnamefont {H.~J.}\ \bibnamefont {Briegel}},\
  }\bibfield  {title} {\bibinfo {title} {Weighted graph states and applications
  to spin chains, lattices and gases},\ }\href
  {https://doi.org/10.1088/0953-4075/40/9/S01} {\bibfield  {journal} {\bibinfo
  {journal} {J. Phys. B: At., Mol. Opt. Phys.}\ }\textbf {\bibinfo {volume}
  {40}},\ \bibinfo {pages} {S1} (\bibinfo {year} {2007})}\BibitemShut {NoStop}%
\bibitem [{\citenamefont {H\"ubener}\ \emph {et~al.}(2009)\citenamefont
  {H\"ubener}, \citenamefont {Kruszynska}, \citenamefont {Hartmann},
  \citenamefont {D\"ur}, \citenamefont {Verstraete}, \citenamefont {Eisert},\
  and\ \citenamefont {Plenio}}]{HuEtAl09}%
  \BibitemOpen
  \bibfield  {author} {\bibinfo {author} {\bibfnamefont {R.}~\bibnamefont
  {H\"ubener}}, \bibinfo {author} {\bibfnamefont {C.}~\bibnamefont
  {Kruszynska}}, \bibinfo {author} {\bibfnamefont {L.}~\bibnamefont
  {Hartmann}}, \bibinfo {author} {\bibfnamefont {W.}~\bibnamefont {D\"ur}},
  \bibinfo {author} {\bibfnamefont {F.}~\bibnamefont {Verstraete}}, \bibinfo
  {author} {\bibfnamefont {J.}~\bibnamefont {Eisert}},\ and\ \bibinfo {author}
  {\bibfnamefont {M.~B.}\ \bibnamefont {Plenio}},\ }\bibfield  {title}
  {\bibinfo {title} {Renormalization algorithm with graph enhancement},\ }\href
  {https://doi.org/10.1103/PhysRevA.79.022317} {\bibfield  {journal} {\bibinfo
  {journal} {Phys. Rev. A}\ }\textbf {\bibinfo {volume} {79}},\ \bibinfo
  {pages} {022317} (\bibinfo {year} {2009})}\BibitemShut {NoStop}%
\bibitem [{\citenamefont {H\"ubener}\ \emph {et~al.}(2011)\citenamefont
  {H\"ubener}, \citenamefont {Kruszynska}, \citenamefont {Hartmann},
  \citenamefont {D\"ur}, \citenamefont {Plenio},\ and\ \citenamefont
  {Eisert}}]{HuEtAl11}%
  \BibitemOpen
  \bibfield  {author} {\bibinfo {author} {\bibfnamefont {R.}~\bibnamefont
  {H\"ubener}}, \bibinfo {author} {\bibfnamefont {C.}~\bibnamefont
  {Kruszynska}}, \bibinfo {author} {\bibfnamefont {L.}~\bibnamefont
  {Hartmann}}, \bibinfo {author} {\bibfnamefont {W.}~\bibnamefont {D\"ur}},
  \bibinfo {author} {\bibfnamefont {M.~B.}\ \bibnamefont {Plenio}},\ and\
  \bibinfo {author} {\bibfnamefont {J.}~\bibnamefont {Eisert}},\ }\bibfield
  {title} {\bibinfo {title} {Tensor network methods with graph enhancement},\
  }\href {https://doi.org/10.1103/PhysRevB.84.125103} {\bibfield  {journal}
  {\bibinfo  {journal} {Phys. Rev. B}\ }\textbf {\bibinfo {volume} {84}},\
  \bibinfo {pages} {125103} (\bibinfo {year} {2011})}\BibitemShut {NoStop}%
\bibitem [{\citenamefont {McLachlan}(1964)}]{Mc64}%
  \BibitemOpen
  \bibfield  {author} {\bibinfo {author} {\bibfnamefont {A.~D.}\ \bibnamefont
  {McLachlan}},\ }\bibfield  {title} {\bibinfo {title} {A variational solution
  of the time-dependent {Schrodinger} equation},\ }\href
  {https://doi.org/10.1080/00268976400100041} {\bibfield  {journal} {\bibinfo
  {journal} {Mol. Phys.}\ }\textbf {\bibinfo {volume} {8}},\ \bibinfo {pages}
  {39} (\bibinfo {year} {1964})}\BibitemShut {NoStop}%
\bibitem [{\citenamefont {\ifmmode \check{C}\else
  \v{C}\fi{}epait\ifmmode~\dot{e}\else \.{e}\fi{}}\ \emph
  {et~al.}(2023)\citenamefont {\ifmmode \check{C}\else
  \v{C}\fi{}epait\ifmmode~\dot{e}\else \.{e}\fi{}}, \citenamefont
  {Polkovnikov}, \citenamefont {Daley},\ and\ \citenamefont
  {Duncan}}]{CeEtAl23}%
  \BibitemOpen
  \bibfield  {author} {\bibinfo {author} {\bibfnamefont {I.}~\bibnamefont
  {\ifmmode \check{C}\else \v{C}\fi{}epait\ifmmode~\dot{e}\else \.{e}\fi{}}},
  \bibinfo {author} {\bibfnamefont {A.}~\bibnamefont {Polkovnikov}}, \bibinfo
  {author} {\bibfnamefont {A.~J.}\ \bibnamefont {Daley}},\ and\ \bibinfo
  {author} {\bibfnamefont {C.~W.}\ \bibnamefont {Duncan}},\ }\bibfield  {title}
  {\bibinfo {title} {{Counterdiabatic Optimized Local Driving}},\ }\href
  {https://doi.org/10.1103/PRXQuantum.4.010312} {\bibfield  {journal} {\bibinfo
   {journal} {PRX Quantum}\ }\textbf {\bibinfo {volume} {4}},\ \bibinfo {pages}
  {010312} (\bibinfo {year} {2023})}\BibitemShut {NoStop}%
\bibitem [{\citenamefont {Lubasch}\ \emph {et~al.}(2016)\citenamefont
  {Lubasch}, \citenamefont {Fuks}, \citenamefont {Appel}, \citenamefont
  {Rubio}, \citenamefont {Cirac},\ and\ \citenamefont {Bañuls}}]{LuEtAl16}%
  \BibitemOpen
  \bibfield  {author} {\bibinfo {author} {\bibfnamefont {M.}~\bibnamefont
  {Lubasch}}, \bibinfo {author} {\bibfnamefont {J.~I.}\ \bibnamefont {Fuks}},
  \bibinfo {author} {\bibfnamefont {H.}~\bibnamefont {Appel}}, \bibinfo
  {author} {\bibfnamefont {A.}~\bibnamefont {Rubio}}, \bibinfo {author}
  {\bibfnamefont {J.~I.}\ \bibnamefont {Cirac}},\ and\ \bibinfo {author}
  {\bibfnamefont {M.~C.}\ \bibnamefont {Bañuls}},\ }\bibfield  {title}
  {\bibinfo {title} {Systematic construction of density functionals based on
  matrix product state computations},\ }\href
  {https://doi.org/10.1088/1367-2630/18/8/083039} {\bibfield  {journal}
  {\bibinfo  {journal} {New J. Phys.}\ }\textbf {\bibinfo {volume} {18}},\
  \bibinfo {pages} {083039} (\bibinfo {year} {2016})}\BibitemShut {NoStop}%
\bibitem [{\citenamefont {Leviatan}\ \emph {et~al.}(2017)\citenamefont
  {Leviatan}, \citenamefont {Pollmann}, \citenamefont {Bardarson},
  \citenamefont {Huse},\ and\ \citenamefont {Altman}}]{LeEtAl17}%
  \BibitemOpen
  \bibfield  {author} {\bibinfo {author} {\bibfnamefont {E.}~\bibnamefont
  {Leviatan}}, \bibinfo {author} {\bibfnamefont {F.}~\bibnamefont {Pollmann}},
  \bibinfo {author} {\bibfnamefont {J.~H.}\ \bibnamefont {Bardarson}}, \bibinfo
  {author} {\bibfnamefont {D.~A.}\ \bibnamefont {Huse}},\ and\ \bibinfo
  {author} {\bibfnamefont {E.}~\bibnamefont {Altman}},\ }\href@noop {}
  {\bibinfo {title} {Quantum thermalization dynamics with {Matrix-Product
  States}}} (\bibinfo {year} {2017}),\ \Eprint
  {https://arxiv.org/abs/1702.08894} {arXiv:1702.08894 [cond-mat.stat-mech]}
  \BibitemShut {NoStop}%
\bibitem [{\citenamefont {White}\ \emph {et~al.}(2018)\citenamefont {White},
  \citenamefont {Zaletel}, \citenamefont {Mong},\ and\ \citenamefont
  {Refael}}]{WhEtAl18}%
  \BibitemOpen
  \bibfield  {author} {\bibinfo {author} {\bibfnamefont {C.~D.}\ \bibnamefont
  {White}}, \bibinfo {author} {\bibfnamefont {M.}~\bibnamefont {Zaletel}},
  \bibinfo {author} {\bibfnamefont {R.~S.~K.}\ \bibnamefont {Mong}},\ and\
  \bibinfo {author} {\bibfnamefont {G.}~\bibnamefont {Refael}},\ }\bibfield
  {title} {\bibinfo {title} {Quantum dynamics of thermalizing systems},\ }\href
  {https://doi.org/10.1103/PhysRevB.97.035127} {\bibfield  {journal} {\bibinfo
  {journal} {Phys. Rev. B}\ }\textbf {\bibinfo {volume} {97}},\ \bibinfo
  {pages} {035127} (\bibinfo {year} {2018})}\BibitemShut {NoStop}%
\bibitem [{\citenamefont {Surace}\ \emph {et~al.}(2019)\citenamefont {Surace},
  \citenamefont {Piani},\ and\ \citenamefont {Tagliacozzo}}]{SuPiTa19}%
  \BibitemOpen
  \bibfield  {author} {\bibinfo {author} {\bibfnamefont {J.}~\bibnamefont
  {Surace}}, \bibinfo {author} {\bibfnamefont {M.}~\bibnamefont {Piani}},\ and\
  \bibinfo {author} {\bibfnamefont {L.}~\bibnamefont {Tagliacozzo}},\
  }\bibfield  {title} {\bibinfo {title} {Simulating the out-of-equilibrium
  dynamics of local observables by trading entanglement for mixture},\ }\href
  {https://doi.org/10.1103/PhysRevB.99.235115} {\bibfield  {journal} {\bibinfo
  {journal} {Phys. Rev. B}\ }\textbf {\bibinfo {volume} {99}},\ \bibinfo
  {pages} {235115} (\bibinfo {year} {2019})}\BibitemShut {NoStop}%
\bibitem [{\citenamefont {Kvorning}\ \emph {et~al.}(2022)\citenamefont
  {Kvorning}, \citenamefont {Herviou},\ and\ \citenamefont
  {Bardarson}}]{KvHeBa22}%
  \BibitemOpen
  \bibfield  {author} {\bibinfo {author} {\bibfnamefont {T.~K.}\ \bibnamefont
  {Kvorning}}, \bibinfo {author} {\bibfnamefont {L.}~\bibnamefont {Herviou}},\
  and\ \bibinfo {author} {\bibfnamefont {J.~H.}\ \bibnamefont {Bardarson}},\
  }\bibfield  {title} {\bibinfo {title} {Time-evolution of local information:
  thermalization dynamics of local observables},\ }\href
  {https://doi.org/10.21468/SciPostPhys.13.4.080} {\bibfield  {journal}
  {\bibinfo  {journal} {SciPost Phys.}\ }\textbf {\bibinfo {volume} {13}},\
  \bibinfo {pages} {080} (\bibinfo {year} {2022})}\BibitemShut {NoStop}%
\bibitem [{\citenamefont {Fr\'{\i}as-P\'erez}\ \emph
  {et~al.}(2024)\citenamefont {Fr\'{\i}as-P\'erez}, \citenamefont
  {Tagliacozzo},\ and\ \citenamefont {Ba\~nuls}}]{FrTaBa24}%
  \BibitemOpen
  \bibfield  {author} {\bibinfo {author} {\bibfnamefont {M.}~\bibnamefont
  {Fr\'{\i}as-P\'erez}}, \bibinfo {author} {\bibfnamefont {L.}~\bibnamefont
  {Tagliacozzo}},\ and\ \bibinfo {author} {\bibfnamefont {M.~C.}\ \bibnamefont
  {Ba\~nuls}},\ }\bibfield  {title} {\bibinfo {title} {{Converting Long-Range
  Entanglement into Mixture: Tensor-Network Approach to Local Equilibration}},\
  }\href {https://doi.org/10.1103/PhysRevLett.132.100402} {\bibfield  {journal}
  {\bibinfo  {journal} {Phys. Rev. Lett.}\ }\textbf {\bibinfo {volume} {132}},\
  \bibinfo {pages} {100402} (\bibinfo {year} {2024})}\BibitemShut {NoStop}%
\bibitem [{\citenamefont {Artiaco}\ \emph {et~al.}(2024)\citenamefont
  {Artiaco}, \citenamefont {Fleckenstein}, \citenamefont {Aceituno~Ch\'avez},
  \citenamefont {Kvorning},\ and\ \citenamefont {Bardarson}}]{ArEtAl24}%
  \BibitemOpen
  \bibfield  {author} {\bibinfo {author} {\bibfnamefont {C.}~\bibnamefont
  {Artiaco}}, \bibinfo {author} {\bibfnamefont {C.}~\bibnamefont
  {Fleckenstein}}, \bibinfo {author} {\bibfnamefont {D.}~\bibnamefont
  {Aceituno~Ch\'avez}}, \bibinfo {author} {\bibfnamefont {T.~K.}\ \bibnamefont
  {Kvorning}},\ and\ \bibinfo {author} {\bibfnamefont {J.~H.}\ \bibnamefont
  {Bardarson}},\ }\bibfield  {title} {\bibinfo {title} {{Efficient Large-Scale
  Many-Body Quantum Dynamics via Local-Information Time Evolution}},\ }\href
  {https://doi.org/10.1103/PRXQuantum.5.020352} {\bibfield  {journal} {\bibinfo
   {journal} {PRX Quantum}\ }\textbf {\bibinfo {volume} {5}},\ \bibinfo {pages}
  {020352} (\bibinfo {year} {2024})}\BibitemShut {NoStop}%
\bibitem [{\citenamefont {Kim}\ \emph {et~al.}(2023)\citenamefont {Kim},
  \citenamefont {Fishman},\ and\ \citenamefont {Sels}}]{KiFiSe23}%
  \BibitemOpen
  \bibfield  {author} {\bibinfo {author} {\bibfnamefont {H.}~\bibnamefont
  {Kim}}, \bibinfo {author} {\bibfnamefont {M.~T.}\ \bibnamefont {Fishman}},\
  and\ \bibinfo {author} {\bibfnamefont {D.}~\bibnamefont {Sels}},\ }\href@noop
  {} {\bibinfo {title} {Variational adiabatic transport of tensor networks}}
  (\bibinfo {year} {2023}),\ \Eprint {https://arxiv.org/abs/2311.00748}
  {arXiv:2311.00748 [quant-ph]} \BibitemShut {NoStop}%
\bibitem [{\citenamefont {Kim}\ \emph {et~al.}(2024)\citenamefont {Kim},
  \citenamefont {Fishman},\ and\ \citenamefont {Sels}}]{KiFiSe24}%
  \BibitemOpen
  \bibfield  {author} {\bibinfo {author} {\bibfnamefont {H.}~\bibnamefont
  {Kim}}, \bibinfo {author} {\bibfnamefont {M.}~\bibnamefont {Fishman}},\ and\
  \bibinfo {author} {\bibfnamefont {D.}~\bibnamefont {Sels}},\ }\bibfield
  {title} {\bibinfo {title} {{Variational Adiabatic Transport of Tensor
  Networks}},\ }\href {https://doi.org/10.1103/PRXQuantum.5.020361} {\bibfield
  {journal} {\bibinfo  {journal} {PRX Quantum}\ }\textbf {\bibinfo {volume}
  {5}},\ \bibinfo {pages} {020361} (\bibinfo {year} {2024})}\BibitemShut
  {NoStop}%
\end{thebibliography}%

\end{document}